\documentclass[sigconf]{acmart}

\AtBeginDocument{%
  }
\usepackage{enumitem}
\usepackage{xcolor}
\usepackage{graphicx} %
\usepackage{duckuments}  
\usepackage{tabularx}

\newcommand{\tool}{CraftTrace}
\newcommand{\q}[1]{\textit{``#1''}}

\usepackage{graphicx}

\newcommand{\modeicon}[1]{%
  \raisebox{-0.16em}{%
    \includegraphics[height=1.05em]{figs/icons/#1.pdf}%
  }%
}

\newcommand{\storyicon}{\modeicon{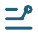}}
\newcommand{\departmentsicon}{\modeicon{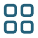}}
\newcommand{\shotsicon}{\modeicon{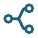}}

\newcommand{\derivationicon}{\modeicon{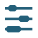}}
\newcommand{\infoicon}{\modeicon{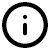}}
\newcommand{\removeicon}{\modeicon{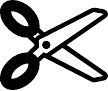}}

\begin{document}

\title{\tool: Unflattening Videos into Malleable, Creation-Inspired Structures for Generative Editing
}


\author{Boyu Li}
\affiliation{%
  \institution{HKUST}
  \city{Hong Kong}
  \country{China}
}

\author{Yuqian Zhou}
\authornote{Corresponding authors.}
\affiliation{%
  \institution{Adobe Research}
  \city{Seattle}
  \state{Washington}
  \country{USA}
}

\author{Duotun Wang}
\affiliation{%
  \institution{HKUST(GZ)}
  \city{Guangzhou}
  \country{China}
}

\author{Ding Li}
\affiliation{%
  \institution{Adobe Research}
  \city{Seattle}
  \state{Washington}
  \country{USA}
}

\author{Zhe Lin}
\affiliation{%
  \institution{Adobe Research}
  \city{Seattle}
  \state{Washington}
  \country{USA}
}

\author{Nanxuan Zhao}
\affiliation{%
  \institution{Adobe Research}
  \city{San Jose}
  \state{California}
  \country{USA}
}

\author{Zeyu Wang}
\affiliation{%
  \institution{HKUST(GZ)}
  \city{Guangzhou}
  \country{China}
}
\affiliation{%
  \institution{HKUST}
  \city{Hong Kong}
  \country{China}
}

\author{Lin-Ping Yuan}
\affiliation{%
  \institution{Nanyang Technological University}
  \country{Singapore}
}

\author{Hongbo Fu}
\authornotemark[1]
\affiliation{%
  \institution{HKUST}
  \city{Hong Kong}
  \country{China}
}



\begin{abstract}
Recent generative video editing models enable video content modification (e.g., changing a character) but target short clips. Extending them to full multi-shot videos requires tedious work to locate relevant content across shots, segment it into clips, craft context-aware editing prompts for each clip, and repeatedly articulate complex editing intent. To address this, we explore an interaction paradigm for editing through underlying video structures (e.g., scripts, scenes, characters, shots, and their relationships). We present CraftTrace, an interactive prototype that transforms a video into a malleable, multilevel structure for generative editing. Users work in task-centric workspaces to modify elements or reshape relationships, while an AI agent translates and propagates changes across the video. A user study and expert review show that this structure helps users understand videos, formulate and refine editing intent, and explore alternatives, supporting rapid prototyping during early-stage exploration and full video post-production.
\end{abstract}




  
\begin{teaserfigure}
  \includegraphics[width=\textwidth]{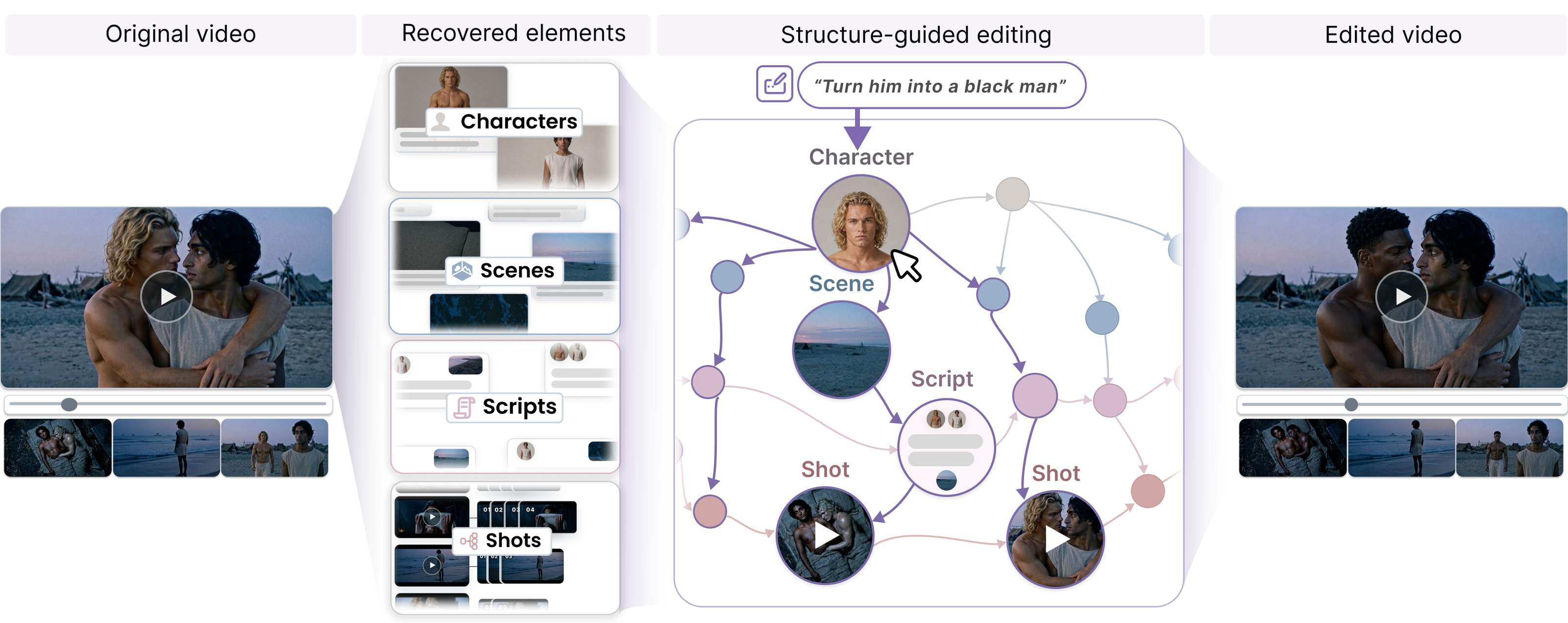}
  \centering
  \caption{\tool~introduces structure-guided generative video editing. Given an input multi-shot full video, the system analyzes its content, extracts key elements, and models their dependencies as a video structure. This structure serves as a scaffold for users to express editing intents on elements or structural relationships, and collaborate with an agent to propagate the resulting changes throughout the video. See the supplementary video for the complete interaction workflow. }
  \label{fig:teaser}
\end{teaserfigure}


\maketitle

\section{Introduction}
Video editing has long centered on transforming existing footage, from cutting and rearranging clips to applying visual effects. Recent advances in generative models are pushing this boundary further, enabling editors to directly modify the visual content within a video~\cite{Wang_2025_CVPR,Mai_2026_CVPR} (e.g., replacing a character or changing the scene). 
However, current generative video editing models accept only short clips as inputs (e.g., up to 30 seconds)~\cite{teamseedance2026seedance}. In practice, novice creators and professional filmmakers often need to edit full multi-shot videos that are minutes or hours long. For example, creators may refine AI-generated videos during post-production without regenerating them from scratch, create entertaining remixes of existing film clips~\cite{constandinides2026para}, or explore alternative creative directions during previsualization and ideation~\cite{runway2026editstudio}. For clarity, we focus on full videos as those exceeding the duration that an editing model supports in a single pass.



Compared to short clips, generative editing of full videos is more difficult in ways that extend beyond merely processing more frames.
(1) Executing edits across a full video requires substantial manual coordination. As videos grow longer, they typically span more shots and scenes, with characters, locations, and events connected across different parts of the video~\cite{CUTTING201469}. An editing target (e.g., a character) may appear only in some clips and under different visual conditions. Creators must therefore repeatedly locate the relevant clips, adapt editing instructions to each clip's context, and maintain consistency across the video.
(2) Editing intents often become more complex in full videos. Creators may first need to understand how different parts of the video fit together before deciding what to change. They may then revise related parts iteratively, using intermediate results to guide further changes. An end-to-end workflow centered on a single prompt and final output offers limited support for this ongoing process of exploration and refinement~\cite{Cao_2025_VideoOrigami}.

To address these challenges, we envision representing a full video through an explicit structure that captures, for example, how its story unfolds and how its characters, settings, and shots are organized. This idea draws inspiration from the video creation process, in which information is distributed across multiple levels, from the overall story and character design to the final shots. Information across these levels is highly interdependent (e.g., each shot is shaped by which characters it includes, where it takes place, and the script). Such a structure can serve as a shared scaffold, providing an AI agent with the context needed to reason about the video and its elements when applying generative edits, while helping users understand the video and iteratively express and refine their editing intent.


\begin{figure*}[ht!]
    \centering
    \includegraphics[width=1\textwidth]{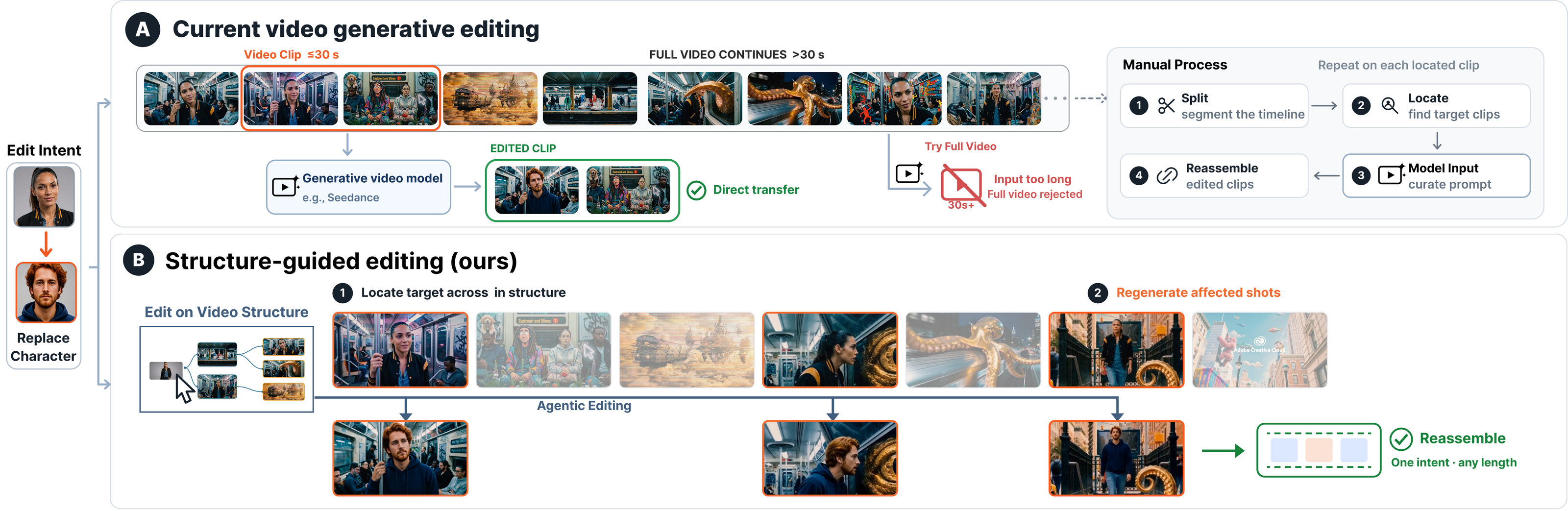}
    \caption{Generative video editing for short clips and full videos. (A) Current models can directly edit short clips (less than 30 s), whereas longer videos require users to manually split, locate, repeatedly edit, and reassemble clips. (B) Our approach provides an interactive video structure through which users express editing intents, with the system automatically propagating across the video.}
    \label{fig:intro}
\end{figure*}
This paper proposes a novel interaction paradigm that uses the structure inferred from video to support interactive agentic video editing. To explore this paradigm, we introduce \tool, a prototype interface for generative video editing. \tool~ transforms a finished video into a multilevel structure that serves as an editable abstraction. Through this abstraction, users can edit not only the clips but also the underlying elements of the video (e.g., its story, scenes, and characters).  However, a single fixed representation may not adequately support the range of editing intents that users bring to a video. 

Drawing on insights from our formative study with six filmmakers, we designed the structure to be malleable in three ways.
First, the representation of the underlying video structure can take four different forms, corresponding to four carefully designed workspaces (i.e., Storyline, Graph, Departments, and Timeline in Fig.~\ref{fig:shapes}), each tailored to a different type of editing task.
Second, changes made through these workspaces can be carried through to the corresponding video content. For example, an edit to a single element can propagate across the structure and ultimately be reflected throughout the full video (Fig.~\ref{fig:intro}). 
Third, users can edit the structure itself by changing the relationships between elements, further expanding the range of possible edits. 
We evaluated \tool~ through a usability study with twelve participants and an expert review with five professional filmmakers. Together, our studies demonstrate that \tool~ can provide an accessible and flexible interface for novice users to post-process existing videos and rapidly explore alternative ideas, while our expert review suggests the potential of the structure-based editing paradigm for professional filmmakers, particularly when integrated with existing AI video creation tools~\cite{TapNow_2026}.

In summary, this paper makes the following contributions:
\begin{itemize}
    \item A design exploration of a novel interactive editing paradigm for generative video editing, using structures reconstructed from existing full videos as a scaffold for expressing and applying editing intent.
    \item \tool, a prototype that instantiates this approach through a malleable structure interface, combining task-centric representations with editable video structures.
    \item Empirical findings from a formative study, a user study, and an expert review that highlight how a structure-based interface helps creators visually explore video, interactively shape and express editing intent, and iteratively refine generative edits.
\end{itemize}

\section{Related Work}
\subsection{Interactive Interfaces for Video Authoring}

Video authoring interfaces have traditionally organized footage along a timeline, making clips and temporal intervals the primary units of interaction. Early HCI research supplemented this representation with metadata~\cite{Mackay_1989,Casares_2002}, detected shot structures~\cite{Ueda_1991,Ueda_1993}, keyframe storyboards~\cite{Girgensohn_2000,Goldman_2006}, hierarchical and multiscale views~\cite{Casares_2002,Ramos_2003}, spatial clip arrangements~\cite{Girgensohn_2003}, semantic segmentation and tagging~\cite{Diakopoulos_2006}, and object-centered navigation~\cite{Dragicevic_2008}. These representations allowed editors to navigate and organize footage at different levels of abstraction. Subsequent systems moved further from temporal coordinates toward representations tailored to particular authoring tasks: storyboards connected planning with production~\cite{Bartindale_2012}; procedural steps structured instructional videos~\cite{Chi_2013,Chi_2021}; captions and plot summaries supported semantic navigation~\cite{Pavel_2015}; transcripts and scripts became direct substrates for editing~\cite{Truong_2016,Leake_2017,Huber_2019,Huh_2023}; and narration~\cite{Xia_2020}, source documents~\cite{Chi_2020,Chi_2022}, or thematic chunks~\cite{Leake_2024} guided the selection and organization of longer-form material. Collectively, these systems demonstrate a shift from editing video solely through time to working with higher-level, task-relevant representations. However, these representations are typically derived from available metadata, annotations, transcripts, scripts, or source documents, and are used primarily to locate, select, arrange, or augment recorded footage rather than to modify its semantic content.

Generative AI has further expanded the role of structure in video authoring. Rather than only describing existing footage, interface structures now also organize the materials and decisions involved in generating it. Language- and agent-based interfaces translate natural-language or multimodal instructions into concrete editing operations~\cite{Tilekbay_2024_ExpressEdit,Wang_2024_LAVE}. Human--AI authoring systems organize source material through narrative framings, scripts, and storyboards~\cite{Wang_2024_ReelFramer}, support semantic clip selection and assembly~\cite{Wang_2024_PodReels}, or align transcripts, timelines, and previews to present alternative edits~\cite{Huh_2025_VideoDiff}. More recent generative environments externalize reusable assets, prompts, and intermediate results through persistent documents~\cite{Liu_2026_Doki}, free-form canvases~\cite{Guo_2026_Protosampling, TapNow_2026}, and spatial layers~\cite{Li_2025_VideoCraft}. VideoOrigami further coordinates scripts, storyboards, canvases, and timelines as synchronized views of the same evolving composition~\cite{Cao_2025_VideoOrigami}, while Vidmento uses the surrounding narrative and visual context to generate new segments around captured footage~\cite{Yeh_2026_Vidmento}. Together, these systems shift structure from passive metadata about a video to an active substrate for generation and human--AI collaboration. However, their structures are generally provided before authoring or constructed as the video is created. And systems that begin with existing media primarily support selecting source material, comparing alternatives~\cite{Huh_2025_VideoDiff}, or generating local extensions~\cite{Yeh_2026_Vidmento}.

Our work addresses a common practical scenario that editing a video when only the finished result is available. This differs fundamentally from authoring a video from scratch, where users define assets and relationships as they work. We adopt a similar infinite-canvas representation but do not require users to manually construct a workflow. Instead, \tool~ decomposes the finished video into editable elements, reconstructs their relationships, and presents them as an interactive graph. Beyond visualization, this recovered structure allows the agent to trace dependencies and coordinate generative edits across the video.

\subsection{Generative Video Editing}

Video editing is no longer limited to cutting, trimming, and arranging clips. Generative models can now modify the visual content of footage while preserving its motion and temporal coherence. Early methods propagated edits through persistent object and background atlases~\cite{Kasten_2021_LayeredAtlases}, composited edit layers~\cite{BarTal_2022_Text2LIVE}, shape-aware deformation fields~\cite{Lee_2023_ShapeAware}, or canonical content fields~\cite{Ouyang_2024_CoDeF}. Diffusion-based approaches substantially expanded this capability, supporting changes to appearance, objects, backgrounds, and style from text prompts~\cite{Wu_2023_TuneAVideo,Qi_2023_FateZero,Ceylan_2023_Pix2Video,Khachatryan_2023_Text2VideoZero,Esser_2023_StructureContent,Yang_2023_Rerender,Liu_2024_VideoP2P,Geyer_2024_TokenFlow}, as well as motion editing~\cite{Molad_2023_Dreamix,Bai_2025_UniEdit,Zhu_2025_FADE}. Subsequent work provides finer control through spatial grounding~\cite{Jeong_2024_GroundAVideo,Yang_2025_VideoGrain}, semantic point correspondences~\cite{Gu_2024_VideoSwap}, edited keyframes~\cite{Fan_2024_Videoshop,Ku_2024_AnyV2V}, direct manipulation~\cite{Deng_2024_DragVideo}, and combinations of source videos, masks, and visual references~\cite{Jiang_2025_VACE}. Instruction-trained models further translate natural-language requests directly into video transformations~\cite{Qin_2023_InstructVid2Vid,Cheng_2024_InsV2V,Zhang_2024_EffiVED,Wu_2025_InsViE,Yu_2025_VEGGIE}, while keyframe, window, and feature propagation methods extend an edit over longer sequences~\cite{Feng_2024_CCEdit,Ma_2024_MaskINT,Zhang_2024_AVID,Liao_2023_LOVECon,Li_2024_VidToMe,Zhang_2025_AdaFlow}.

Recent interfaces and agents move beyond rendering to help users specify and carry out editing operations. ExpressEdit grounds temporal, spatial, and operational intent from language and sketches~\cite{Tilekbay_2024_ExpressEdit}; LAVE and AssistEditor plan actions or operate editing tools from high-level requests~\cite{Wang_2024_LAVE,Gao_2024_AssistEditor}; and VideoDiff presents aligned alternatives for rough cuts, B-roll, and effects~\cite{Huh_2025_VideoDiff}. TeaserGen retrieves and assembles excerpts from long documentaries~\cite{Xu_2025_TeaserGen}, while EditDuet coordinates editor and critic agents for non-linear editing~\cite{SandovalCastaneda_2025_EditDuet}. More recent systems prepare model-ready editing conditions~\cite{Yu_2026_Aurora} or coordinate specialized agents across long-form production workflows~\cite{Zhou_2026_VideoAgent,Zhao_2026_CutClaw,Yan_2026_Crayotter}.

Despite these advances, most generative editors work on a clip that has already been selected and apply an edit across its frames~\cite{Geyer_2024_TokenFlow,Ku_2024_AnyV2V,Jiang_2025_VACE,Zhang_2025_AdaFlow}. AI agents can process more footage, but mainly retrieve, arrange, or trim existing clips~\cite{Xu_2025_TeaserGen,SandovalCastaneda_2025_EditDuet,Zhao_2026_CutClaw}. Applying generative edits to a finished film presents a different challenge: the same character, scene, or narrative element may appear across many nonadjacent shots. \tool~ therefore recovers a semantic structure from the finished video and uses it to identify the affected shots, adapt the edit to each shot, select appropriate visual references, and coordinate the generative models.

\subsection{Representing Creative Processes}
Our motivation for building a semantic structure for video editing comes from the rich decisions embedded in the video creation process. We therefore first review how prior work has represented, recorded, and used creative processes.

Early work on design rationale therefore represented the issues, alternatives, and arguments behind an artifact~\cite{Conklin_1988_gIBIS,MacLean_1989_DesignRationale}. Subsequent history systems captured intermediate states and connected parts of an artifact to the actions and workflows that created them~\cite{Klemmer_2002_DesignHistory,Grossman_2010_Chronicle}, turning history from an undo record into a resource for understanding and revising prior work. Creative processes, however, rarely follow a single path. Later systems therefore represented causal relationships between actions~\cite{Nancel_2014_Causality}, retained simultaneous variations~\cite{Terry_2004_Variation}, supported parallel exploration~\cite{Hartmann_2008_Juxtapose,Dow_2011_ParallelPrototyping}, organized alternatives through branching and merging~\cite{Zaman_2015_GEMNI}, and treated versions as materials for reflection and reuse~\cite{Sterman_2022_CreativeVersionControl}. Together, these works show that a process representation is valuable not simply because it records what happened, but because it makes the alternatives, dependencies, and rationale behind an outcome available for further creative work.

Generative AI makes such representations both more important and more difficult to maintain. A generated artifact may depend on prompt wording, examples, references, model choices, and repeated evaluations~\cite{Subramonyam_2025_PrototypingPrompts}, yet these decisions are easily hidden behind the final output. Recent systems therefore turn prompts and model configurations into persistent, manipulable objects~\cite{Angert_2023_Spellburst,Kim_2023_Cells,Riche_2025_AIInstruments}, visualize how ideas and alternatives evolve~\cite{Shen_2025_IdeationWeb,Peng_2026_DesignTrace}, and connect artifact states to the discussions that motivated them~\cite{Li_2025_DesignMemo}. Complementary work automatically summarizes relationships among creative actions~\cite{Smith_2025_FuzzyLinkography}, shows that artifact and command histories alone omit goals and rationale~\cite{Cheng_2026_LostTranslation}, and examines how creative traces and provenance support reflection, interpretation, and reuse~\cite{Moruzzi_2025_ContentAuthenticities,Hammad_2026_TracingCreativity}.

Prior systems primarily capture creative traces as creation unfolds, using them to preserve, inspect, or revisit the original process. \tool~ starts from a different condition: only the finished video is available. We infer a plausible, creation-inspired structure not to reconstruct how the video was actually made, but to organize its elements and dependencies into an editable scaffold. Thus, prior work uses traces to explain past creation, whereas we use the creation process as a lens for structuring future edits.

\section{Formative Study}

To explore the structure-based interaction paradigm, we conducted two formative studies. We first interviewed video creators about how they approach full video edits that span multiple shots and what support they need. We then built a tech probe that exposed the recorded agentic video creation process as a graph structure and allowed participants to inspect its intermediate elements. 

\subsection{Interview Study}

\subsubsection{Participants and Procedure.}
We recruited six filmmakers (I1-I6) with experience creating and editing videos. All had previously used generative image or video tools. We focused on generative, content-level edits that span multiple shots (e.g., replacing a character, changing a setting, or introducing new story content), rather than conventional operations such as trimming or rearranging clips. Such edits require filmmakers to coordinate shared elements and creative decisions across the video.
Participants first described a recent video project and walked us through their current workflow, focusing on how they handled changes that affected multiple shots and where difficulties arose. We then discussed when and why they might use generative editing in video production. The interview then turned to whether the video structure could serve as a medium for such edits. Participants reflected on how intermediate elements in the creation process would be useful, and how they might be organized for editing. 
Each interview lasted approximately 45 minutes and was recorded and transcribed for analysis.

\subsubsection{Current Practices and Generative Editing Needs}
Participants described cross-shot generative editing as a broader creative practice in both professional and personal contexts, including \q{previsualizing creative alternatives}(I1, I4), \q{studying and reinterpreting existing works}(I1, I5), \q{creating playful personalized versions}(I6), and \q{localizing AI short dramas for different audiences}(I3). These edits typically target more high-level elements (e.g., character, story) rather than individual clips. Current workflows require creators to locate and modify the affected shots one by one, repeatedly specify similar prompts, and manually check consistency across the video. 

Participants also emphasized that the final footage \q{[only shows the result]} (I1) and hides many of the decisions needed to revise it. In practice, they often had to \q{[go back to the script and character designs]} (I5) to understand how a shot was constructed and identify where the same decisions recurred. As I2 noted,\q{The structure provides an alternative way to index the video. While a traditional timeline organizes content by time, a creation trace could organizes it by meaning and dependency.} Keeping these intermediate elements and their relationships accessible could be a valuable editing scaffold.

\subsection{Tech-Probe Study}
\label{sec:tech_probe}
Building on the interview findings, we designed a tech probe to explore whether making a video’s creation process visible could help creators understand and coordinate edits across multiple shots. Reconstructing this process from a finished video would introduce a separate technical challenge and make its value harder to assess. We therefore generated a video through an agentic workflow and constructed the graph structure directly from its recorded creation trace. Using this known trace kept the study focused on whether the creation process could serve as an editing scaffold, before we invested substantial effort in recovering it from existing videos.

\subsubsection{Participants and Procedure.}
To examine the tech probe, we recruited six filmmakers (P1--P6) with experience creating and editing multi-shot videos. All had previously used generative image or video tools. We prepared five videos using Codex with a custom video-creation skill. The skill recorded each video’s story events, characters, scenes, shot scripts, and generated clips, which we converted into a node-link graph showing how these elements came together in the final video.
After a brief walkthrough, participants watched each video and explored its process graph. They thought aloud while tracing relationships across the graph, identifying elements they might modify, and describing how those changes should affect the rest of the video. We concluded with a semi-structured interview about the representation’s usefulness, desired editing interactions, and possible improvements. Each session lasted approximately 20 minutes.

\subsubsection{Tech-Probe Finding}
We found that representations of the creation process can scaffold users’ understanding of an existing video and support its subsequent modification with generative models. Building on this potential, we sought to design a structured editing interface derived from the creation process. However, our tech probe revealed three challenges in transforming process representations into an effective editing medium.

\textbf{Mismatch between fixed structure and task-specific editing needs (C1).} A video creation process spans multiple levels (e.g., narrative development, concept design, and shot structure). However, participants did not consider all of these levels equally relevant to every editing task. When asked to envision how they might use a process representation for editing, they described a wide range of goals that required different information and levels of detail. Some participants wanted to explore the overall narrative without engaging with low-level production details. As P1 explained, \q{I don’t want to get into shot-level details. I just want to see how different storylines might unfold.} In contrast, P5 wanted focused support for character design: \q{When I focus on character development, I do not want to be distracted by unrelated details. I would prefer a dedicated workspace that provides more context for that task.} Together, these responses reveal a mismatch between a fixed, comprehensive representation of the creation process and the task-specific ways in which participants wanted to engage with it. 

\textbf{Burden of realizing global edits through repeated local modifications (C2).} Although our tech probe made it straightforward to edit individual nodes on canvas, participants (5/6) expected a single edit to be reflected across all relevant parts of the video. As P4 noted, \q{When I modify a character, I want the system to automatically determine when that character appears and update only those shots.} 
In addition, participants (2/6) also emphasized the potential complexity of this automated process, particularly when formulating prompts for visual editing in each local context. As P6, who frequently used generative editing tools, explained, \q{When editing an image or video, the prompt has to be specific to the task—based on what the current image looks like, what I want it to become, and what should remain unchanged. Even if the overall change is simple, different shots still need different instructions because their content is different.} Thus, an edit may need to be decomposed into many local editing tasks, each of which must be dynamically constructed and coordinated based on its specific context while preserving the overall editing intent.

\textbf{Difficulty in expressing relational editing intents (C3).} When interacting with the process map, half of the participants (3/6) wanted to do more than modify individual nodes; they also wanted to reshape how nodes were connected. For example, P2 explained, \q{I want to remove the connection between Character 1 and this shot so that the character is removed from the shot.} Similarly, P5 noted, \q{It would be interesting if I could reconnect parts of the process and have the agent understand the intention behind the new connection.} These responses suggest that participants did not view the process map as a fixed workflow, but as a malleable structure in which connections themselves could express editing intentions. This introduces the challenge of interpreting what an added, removed, or modified connection means and translating that change into the resulting video.


\subsection{Design Goal: Malleable Process as Scaffold}
To address these challenges, we define a malleable process as a structured editing workspace derived from the video creation process that users can flexibly access, navigate, and modify to express diverse editing intents and transform the underlying video with:

\begin{itemize}[left=0pt]

    \item \textbf{DG1. Adapt the video structure to users’ current editing tasks.} Rather than presenting every user with the same process interface regardless of their editing goal, the interface should be carefully designed to accommodate diverse needs, drawing on the envisioned use cases identified in the formative study.
    
    \item \textbf{DG2. Propagate a single edit across the video.} Rather than requiring users to modify elements one by one, the system should automatically infer their editing intent, propagate it across dependent stages, and adaptively translate the global intent into a set of targeted local editing tasks.
    
    \item \textbf{DG3. Make the video structure itself editable.} Rather than treating connections between process nodes as fixed, we allow users to add, remove, or modify these connections, with the resulting changes reflected in the final video semantically.
\end{itemize}

\begin{figure*}[h]
  \includegraphics[width=1\textwidth]{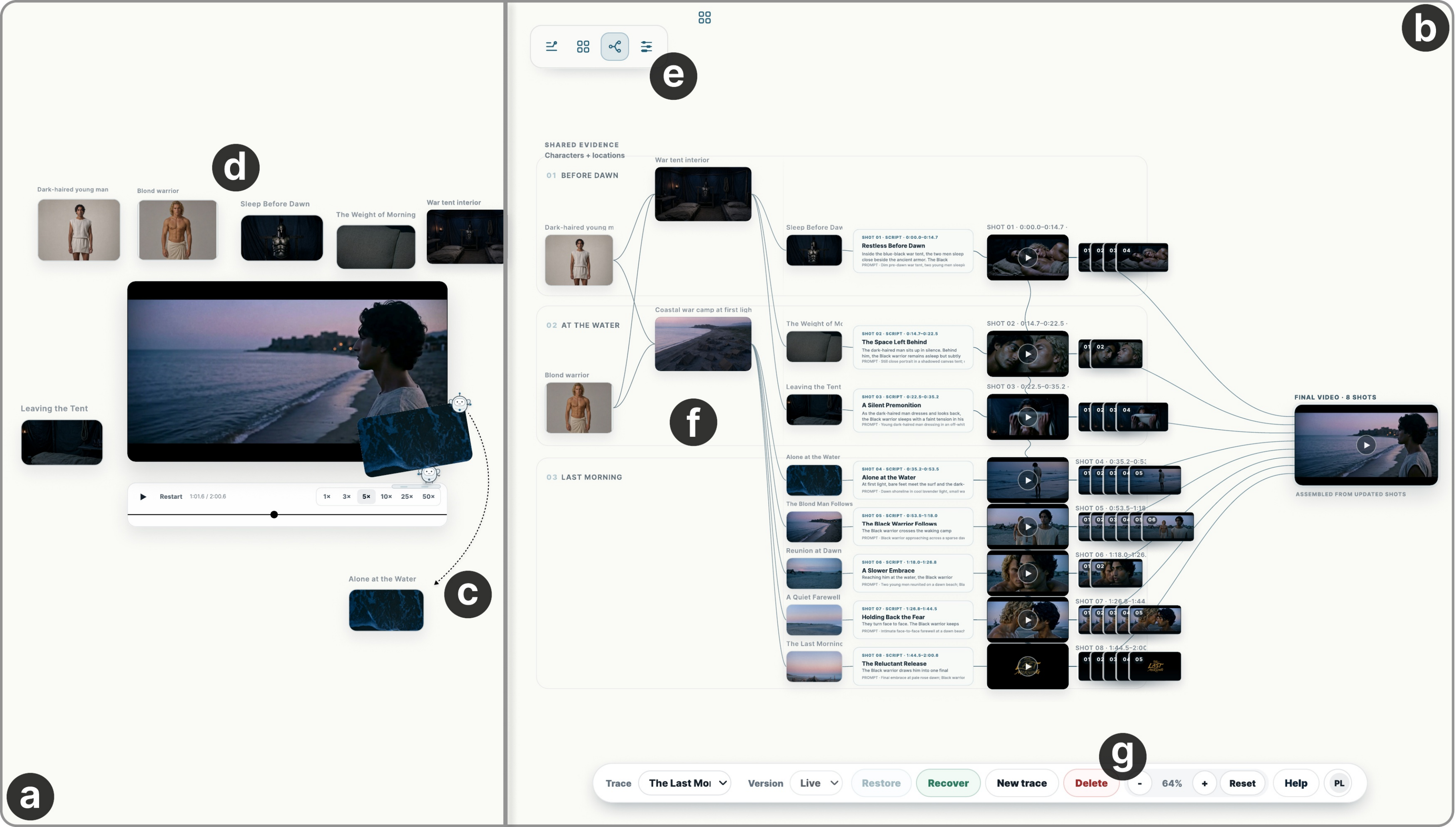}
  \centering
  \caption{The \tool~interface. (a) Video analysis view and (b) structure-based editing interface. (c) The agent first extracts video elements, such as (d) characters and scenes, and establishes their identities and dependencies. (e) The interface organizes the extracted structure into four workspaces, with the Graph workspace shown as an example. (f) The graph visualizes the relationships between the extracted elements and video clips. (g) The toolbar provides basic functions, such as version control, canvas manipulation, and project management.}
  \label{fig:interface}
\end{figure*}
\section{\tool}

In this section, we explore how \tool~ is grounded in a video’s underlying structure to support high-level generative editing(Fig.~\ref{fig:interface}). Our approach consists of two system stages.

The first stage, \textbf{reconstructing the creation process structure}, infers a plausible process representation from the rendered video, as an existing video does not expose how it was created (implementation in Sec.~\ref{sec:reconstruct}). This representation models narrative events, characters, scenes, scripts, shots, and video segments as interconnected elements. Based on the editing needs identified in our formative study, \tool~ presents this shared structure through four task-centric workspaces: Storyline, Graph, Departments, and Timeline. Each workspace foregrounds a different aspect of the video structure and provides a corresponding entry point for understanding it (Sec.~\ref{sec:reshape}).

The second stage, \textbf{agentic video editing based on the structure}, allows users to express editing intents through these workspaces (implementation in Sec.~\ref{sec:guide}). When a user modifies an individual element, the agent traces its dependencies, coordinates changes across the affected structure, and carries the results through to the final video (Sec.~\ref{sec:rewrite}). Some editing intents, however, require changing not only the content of individual elements but also the relationships among them. For these cases, users can modify the structure itself, and the agent interprets these structural changes as semantic guidance for reorganizing the affected content (Sec.~\ref{sec:relink}).


\par\smallskip
\noindent
\begingroup
\setlength{\fboxsep}{5pt} 
\colorbox{black!5}{%
  \parbox{\dimexpr\columnwidth-2\fboxsep\relax}{%
    \hspace*{\parindent}%
    \textit{The following sections use Gabriel's experience as a
    \textbf{running example} to illustrate the system design. Gabriel
    encounters a video online and wants to modify its content. We follow
    how he uses the malleable process structure reconstructed by the system
    as a scaffold to iteratively edit the video through intuitive interactions.}
  }%
}
\endgroup
\par\smallskip
\subsection{Task-Centric Visualization of Video Structure}
\label{sec:reshape}

\par\vspace{1pt}
\noindent
\begingroup
\setlength{\fboxsep}{5pt} 
\colorbox{black!5}{%
  \parbox{\dimexpr\columnwidth-2\fboxsep\relax}{%
    \hspace*{\parindent}%
    \textit{Gabriel first imports the video into the system. Once the system finishes analyzing it, he is presented with a graph showing how characters, scenes, and shots are connected to form the final video. He can switch between different representations of the same underlying process, including a conventional timeline, a step-by-step storyline, and detailed specifications of characters, scenes, and other production elements.}
  }%
}
\endgroup
\par\smallskip

\begin{figure*}[h]
  \includegraphics[width=0.9\textwidth]{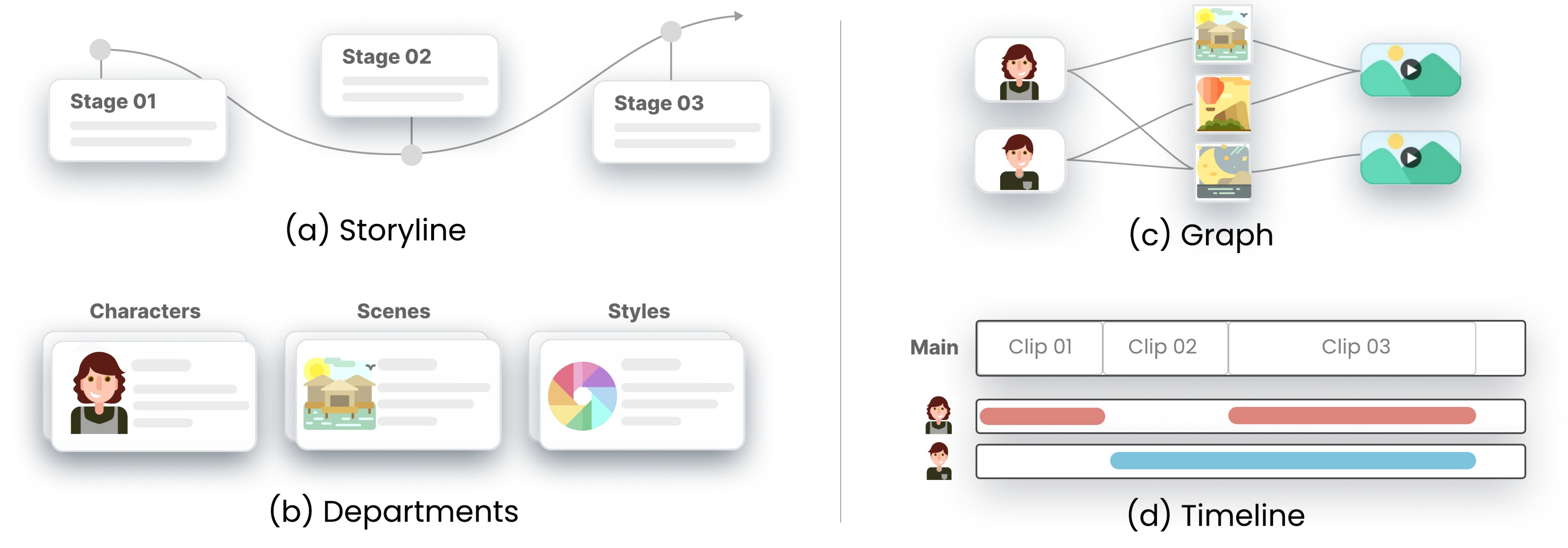}
  \centering
  \caption{Four workspaces of \tool. }
  \label{fig:shapes}
\end{figure*}
To support diverse user tasks (\textbf{DG1}), we conducted thematic coding of participants' envisioned tasks from the formative study, identifying four editing focuses. Inspired by established interface representations~\cite{Cao_2025_VideoOrigami}, we translated these focuses into four task-oriented workspaces (Fig.~\ref{fig:shapes}) that exposing different aspects of the video’s underlying structure: 

\textit{\storyicon~\textbf{Storyline.}}  A video is not merely a sequence of visual content, but also conveys a story. While most video editing research focuses on the visual frames~\cite {berniniteam2026bernini,xiao2026joyaivideoedit},
we design the Storyline workspace that represents the video in terms of its story. As shown in Fig.~\ref{fig:shapes}-a, the workspace presents the narrative structure inferred from the video as a sequence of temporally ordered nodes, each summarizing a major event or plot segment. Our current prototype focuses on linear narratives to illustrate the core interactions, but this representation could be extended to support nonlinear narrative structures further.

\textit{\shotsicon~\textbf{Graph.}}
The interface organizes the extracted elements into a multilevel dependency graph linking narrative events, scenes, characters, and shots (Fig.~\ref{fig:shapes}-c). Users can hover over a node to highlight its connections and trace its relationships across the video. For example, hovering over a character reveals the scenes in which the character appears. Users can also select a node and click the Info button (\infoicon) to inspect its details, such as the inferred generation prompt for a video clip.

\textit{\departmentsicon~\textbf{Departments.}}
Traditional filmmaking distributes creative decisions across specialized departments. For example, casting and costume teams shape characters, and production designers define scene settings. Although handled separately during production, these elements are integrated in the video and collectively shape its visual presentation. \tool~therefore organizes the recovered elements by production department (Fig.~\ref{fig:shapes}-b), allowing users to inspect the design attributes of each element, such as a character's age, clothing, and hair color.

\begin{figure*}[h]
  \includegraphics[width=1\textwidth]{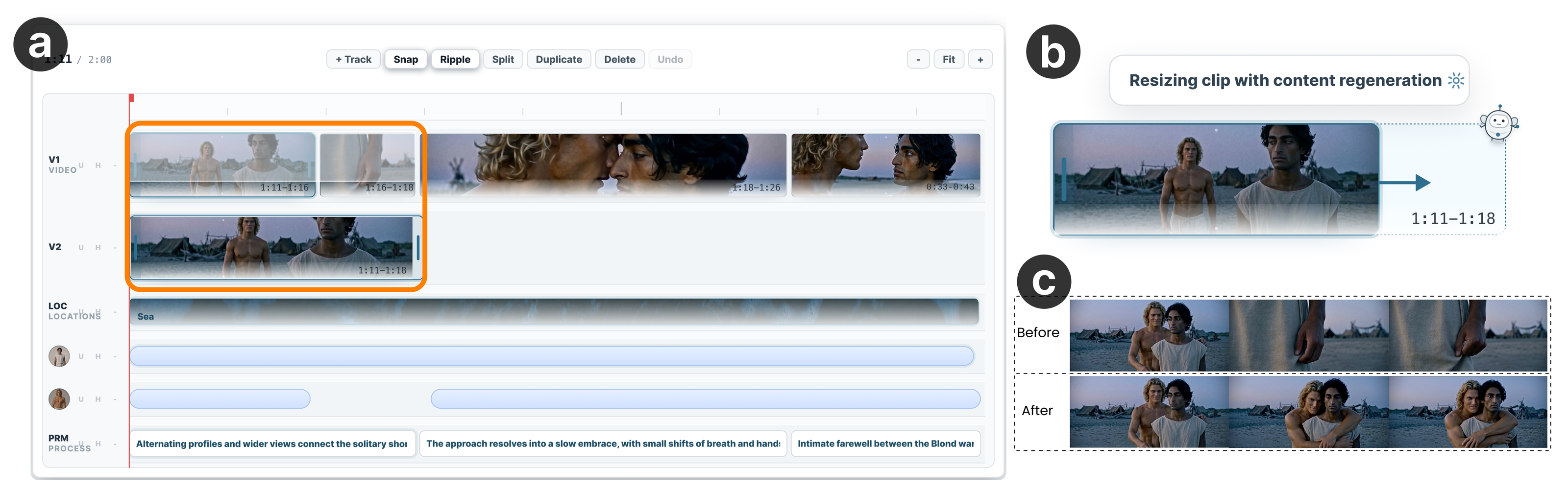}
  \centering
  \caption{\tool~ supports users in expressing generative editing intents through direct manipulation in the \derivationicon~Timeline workspace. For example, (a) a user deletes a clip and extends the preceding one to fill the gap; (b) the agent interprets this action as a request to continue the preceding clip; and (c) after receiving the user’s confirmation, it generates an extension based on that clip, replacing the deleted segment with a seamless continuation.}
  \label{fig:timeline}
\end{figure*}

\textit{\derivationicon~\textbf{Timeline.}} Conventional timelines support temporal navigation and clip-level operations. CraftTrace extends this representation with semantic tracks for scripts, characters, scenes, and shots (Fig.~\ref{fig:shapes}-d, Fig.~\ref{fig:timeline}-a), allowing users to inspect when each element appears and how it contributes to individual shots.


\subsection{Editing through the Structure}
\label{sec:rewrite}
\par\vspace{1pt}
\noindent
\begingroup
\setlength{\fboxsep}{5pt} 
\colorbox{black!5}{%
  \parbox{\dimexpr\columnwidth-2\fboxsep\relax}{%
    \hspace*{\parindent}%
    \textit{After exploring these workspaces for approximately ten minutes, Gabriel develops a clear understanding of the video. He then begins using these workspace to modify the video and explore alternative creative directions.}
  }%
}
\endgroup
\par\smallskip

Current generative authoring interfaces often require users to edit elements individually, such as selecting an image or video node and applying an AI edit~\cite{Guo_2026_Protosampling}. This places the burden on users to identify and update every element affected by a change. \tool~automates this coordination by tracing dependencies between elements, enabling an agentic workflow grounded in the recovered structure (\textbf{DG2}). For example, when a user modifies a character, the agent identifies the shots in which the character appears and propagates the change to them. We integrate this workflow into the four workspaces introduced above, each providing a context for a different type of video editing task.

\begin{figure*}[h]
  \includegraphics[width=1\textwidth]{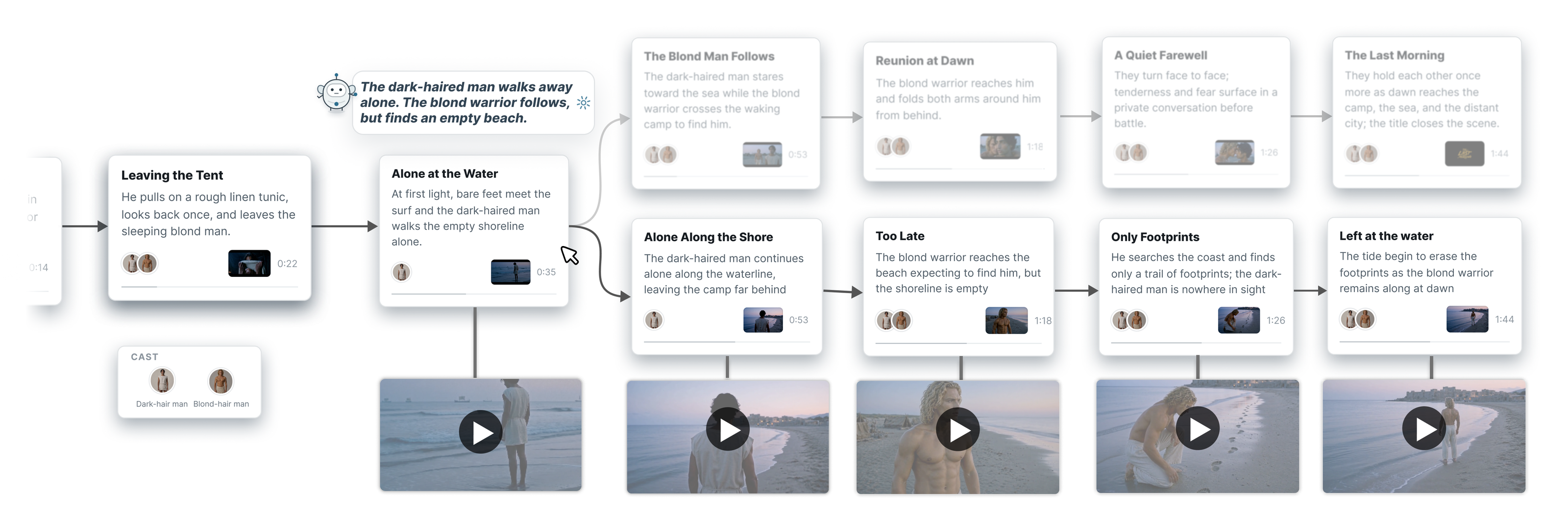}
  \centering
  \caption{\tool~supports editing a video story with generative editing. In the \storyicon~Storyline workspace, each event corresponds to a video clip. The user could modify the narrative event following the node \q{Alone at the Water}. The interface generates a new event (bottom right) and its corresponding video, replacing the original event and clip (top right).}
  \label{fig:story}
\end{figure*}

\subsubsection{Editing in \storyicon~Storyline: Exploring Narrative Alternatives}

\mbox{}\par
\vspace{1pt}
\noindent
\begingroup
\setlength{\fboxsep}{5pt} 
\colorbox{black!5}{%
  \parbox{\dimexpr\columnwidth-2\fboxsep\relax}{%
    \hspace*{\parindent}%
    \textit{Gabriel wants to brainstorm alternative directions for the storyline. In the \storyicon~Storyline workspace, he selects the node from which he wants the story to diverge and describes his desired change. \tool~ then replans the subsequent events and progressively generates the revised video clips.}
  }%
}
\endgroup
\par\smallskip

Modifying a video's storyline often requires users to rewrite prompts and regenerate multiple segments, making it difficult to explore alternative narratives during ideation. To address this, the \storyicon~Storyline workspace allows users to modify the narrative directly through the story structure presented on the canvas. As illustrated in Fig.~\ref{fig:story}, users can either describe the desired change in a simple prompt and let the agent revise the story or edit the events manually. Once users approve the revision, \tool~automatically selects the appropriate character and scene references, generates the corresponding video segments, and replaces the original segments.

\begin{figure*}[h]
\includegraphics[width=1\textwidth]{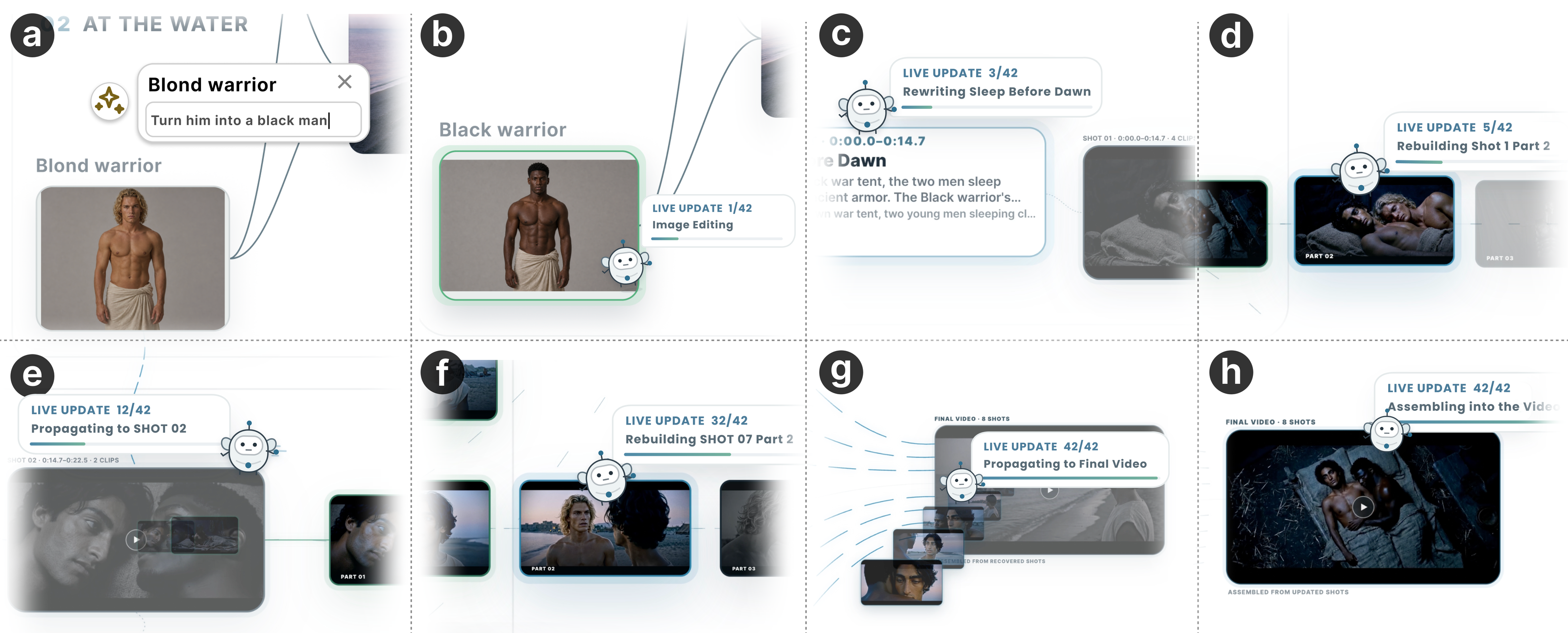}
  \centering
  \caption{\tool~propagates a single node-level edit throughout the video structure automatically. In the \shotsicon~Graph workspace  (full interface is shown in Fig.~\ref{fig:interface}-right), the user (a) selects a character and describes the desired change. The agent then (b) updates the character and traces its dependencies, (c) revises the related scripts, (d) regenerates the affected shots, and (e) reflects these changes in the sequence. Throughout this process, it (f) determines which elements to update and which to preserve. Once (g) the revised clips are ready, \tool~ (h) assembles them into the updated video.}
  \label{fig:graph}
\end{figure*}

\subsubsection{Editing in \shotsicon~Graph: Propagating Edits through Dependencies}
\mbox{}\par
\vspace{1pt}
\noindent
\begingroup
\setlength{\fboxsep}{5pt} 
\colorbox{black!5}{%
  \parbox{\dimexpr\columnwidth-2\fboxsep\relax}{%
    \hspace*{\parindent}%
    \textit{Gabriel next wants to replace a character throughout the video. In the \shotsicon~Graph workspace, he selects the character and enters a prompt. A small agent springs to life and traverses the graph, carrying the change into every affected node.}
  }%
}
\endgroup

\par\smallskip
Although we could design the agent to execute an edit end to end, doing so would give users limited visibility into how the edit is interpreted and carried out across the video. We therefore use the \shotsicon~Graph workspace to externalize this process. When a user edits an element node, \tool~traces its relationships to identify the affected subgraph and determine which elements and shots should be updated or preserved (Fig.~\ref{fig:graph}). The interface visualizes the agent's progress as the edit propagates through these dependencies. Users can inspect the process, restrict its scope, or set breakpoints before particular changes are executed.

\begin{figure*}[h]
  \includegraphics[width=1\textwidth]{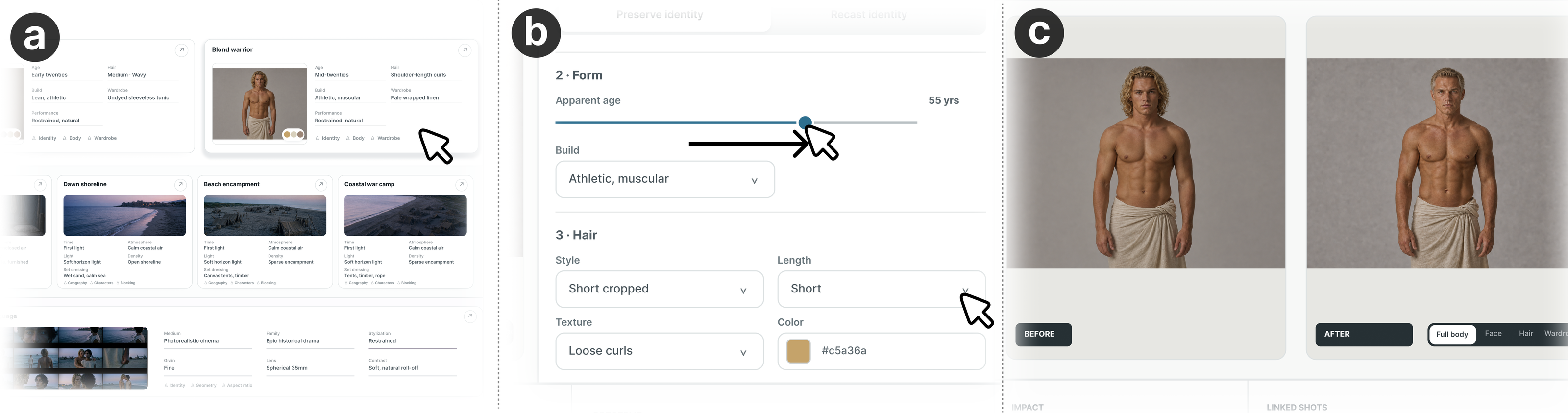}
  \centering
  \caption{\tool~ supports detailed attribute editing of video elements in the \departmentsicon~Departments workspace. (a) Elements such as characters and scenes are organized in sheets. (b) Users can open an element to edit its on-demand attributes. (c) Through iterative preview and adjustment, users can refine the edited identity and then apply the result across the entire video.}
  \label{fig:department}
\end{figure*}


\subsubsection{Editing in \departmentsicon~Departments: Refining Production Elements}
\mbox{}\par
\vspace{1pt}
\noindent
\begingroup
\setlength{\fboxsep}{5pt} 
\colorbox{black!5}{%
  \parbox{\dimexpr\columnwidth-2\fboxsep\relax}{%
    \hspace*{\parindent}%
    \textit{Gabriel finds prompting in the graph too coarse and switches to the \departmentsicon~Departments workspace for finer control. He iteratively adjusts the character’s attributes, reviews the generated results, and applies the final design to the video.}
  }%
}
\endgroup
\par\smallskip

Video creators often refine individual elements in detail to achieve the intended visual result. To reflect this practice, the \departmentsicon~Departments workspace makes the attributes of recovered elements explicit and editable. For each selected element, \tool~identifies relevant attributes and generates controls for adjusting them on demand. Users can iteratively refine the element and preview alternative designs before selecting a desired version. Once confirmed, \tool~applies the updated element to the relevant shots through generative video editing.

\subsubsection{ Editing in \derivationicon~Timeline: Expressing Generative Intent through Direct Manipulation}
\mbox{}\par
\vspace{1pt}
\noindent
\begingroup
\setlength{\fboxsep}{5pt} 
\colorbox{black!5}{%
  \parbox{\dimexpr\columnwidth-2\fboxsep\relax}{%
    \hspace*{\parindent}%
    \textit {Finally, Gabriel turns to clip-level editing in the \derivationicon~Timeline workspace. He removes a close-up shot of a hand, leaving a gap in the sequence. To fill it, he drags the end of the preceding shot across the empty interval. The agent then appears and suggests generating a continuation of the preceding shot. Gabriel accepts, and the agent fills the gap with the generated extended shot.}
  }%
}
\endgroup
\par\smallskip

In conventional timelines, deleting, moving, or resizing a clip only changes its position or duration. Generative edits, however, are usually specified through separate text prompts. \tool~ connects these two forms of interaction by interpreting timeline operations as editing intent. For example, extending a video clip beyond its original endpoint indicates both that the scene should continue and how long the generated continuation should be. 
Since these operations can be ambiguous, the agent first presents its interpretation of the user’s action. Users can then either confirm the interpretation or provide a short prompt to clarify their intent before any changes are made. 
\begin{figure*}[h!]
  \includegraphics[width=1\textwidth]{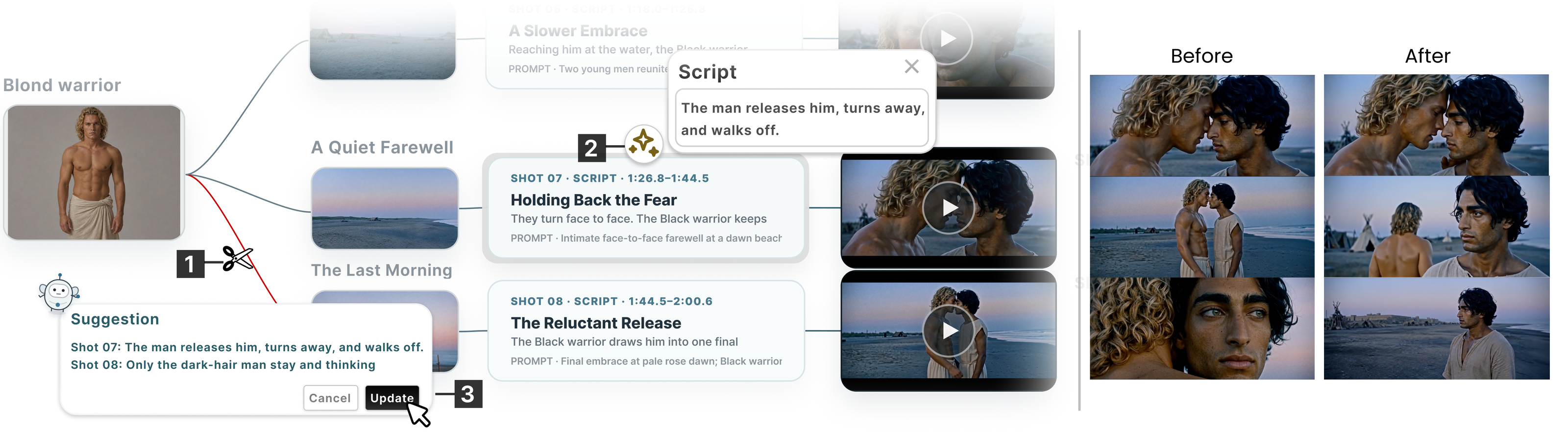}
  \centering
  \caption{\tool~ supports structural editing in the \shotsicon~Graph workspace. Users can enter connection-editing mode and modify relationships between video elements. For example, (1) the user removes the connection between a character and a following shot, indicating that the character should no longer appear in that shot; (2) the user further revises the preceding script to make the change narratively coherent; and (3) these operations are first staged for review. Once confirmed, the agent infers corresponding editing actions from the staged operations, and the user can click \texttt{Update} to apply them. (Right) The affected video content is regenerated accordingly.  }
  \label{fig:relink}
\end{figure*}

\subsection{Reshape the Video Structure}
\label{sec:relink}

\par\vspace{1pt}
\noindent
\begingroup
\setlength{\fboxsep}{5pt}
\colorbox{black!5}{%
  \parbox{\dimexpr\columnwidth-2\fboxsep\relax}{%
    \hspace*{\parindent}%
    \textit{Gabriel wants a character to leave at the end of a scene, but the current structure keeps the character connected to the following shot. He switches to \textsc{Relink} mode, removes this connection, and provides a short prompt for revising the script. The agent infers his intent from these actions and proposes an updated scene. After Gabriel confirms the proposal, the agent regenerates the affected clips to show the character leaving.}
  }%
}
\endgroup
\par\smallskip

The interactions described above use a fixed structure as a scaffold for editing. While this works well for many tasks, some editing intents require changing the structure itself (e.g., removing a character from a shot or introducing a new scene). Such changes cannot be fully expressed by modifying a single node, and they require changing how nodes are connected (\textbf{DG3}). To support such edits, we make the recovered structure itself malleable, allowing users to add, remove, or reconnect relationships to reshape the structure.

The structure can be edited in the \shotsicon~Graph workspace, where the video structure and the relationships among its elements are visible together. Users can click the \textsc{Relink} button in the upper-left corner to reveal a set of tools for editing these relationships. They can select the \removeicon~remove tool to remove an existing edge, or drag from the circular handle on a node to another node to create or redirect a connection.

Through these structure editing operations, users can express how they want the video to change. Rather than interpreting each operation in isolation, the agent considers multiple related operations together within the context of the current graph to infer the user's broader editing intent.
For example, in Fig.~\ref{fig:relink}, the user removes the connection between a character and the final shot and revises the script of the preceding shot. By considering these operations together, the agent interprets that the character should leave during the preceding shot and no longer appear in the final shot. It presents this interpretation and the proposed updates to the user. Once confirmed, the agent updates the affected nodes in sequence and regenerates the corresponding video segments.

\par\vspace{1pt}
\noindent
\begingroup
\setlength{\fboxsep}{5pt}
\colorbox{black!5}{%
  \parbox{\dimexpr\columnwidth-2\fboxsep\relax}{%
    \hspace*{\parindent}%
    \textit{Editing operations are first collected in a staging area, where Gabriel can review or undo them before committing the changes. Once confirmed, the agent interprets the staged operations as editing intent and executes the corresponding tasks. After processing is complete, Gabriel can download the revised video. He can also open the version history to restore any earlier version.}
  }%
}
\endgroup
\par\smallskip
\section{Implementation}

\subsection{Reconstruction of Video Structure}
\label{sec:reconstruct}
\begin{figure*}[h!]
    \centering
    \includegraphics[width=1\textwidth]{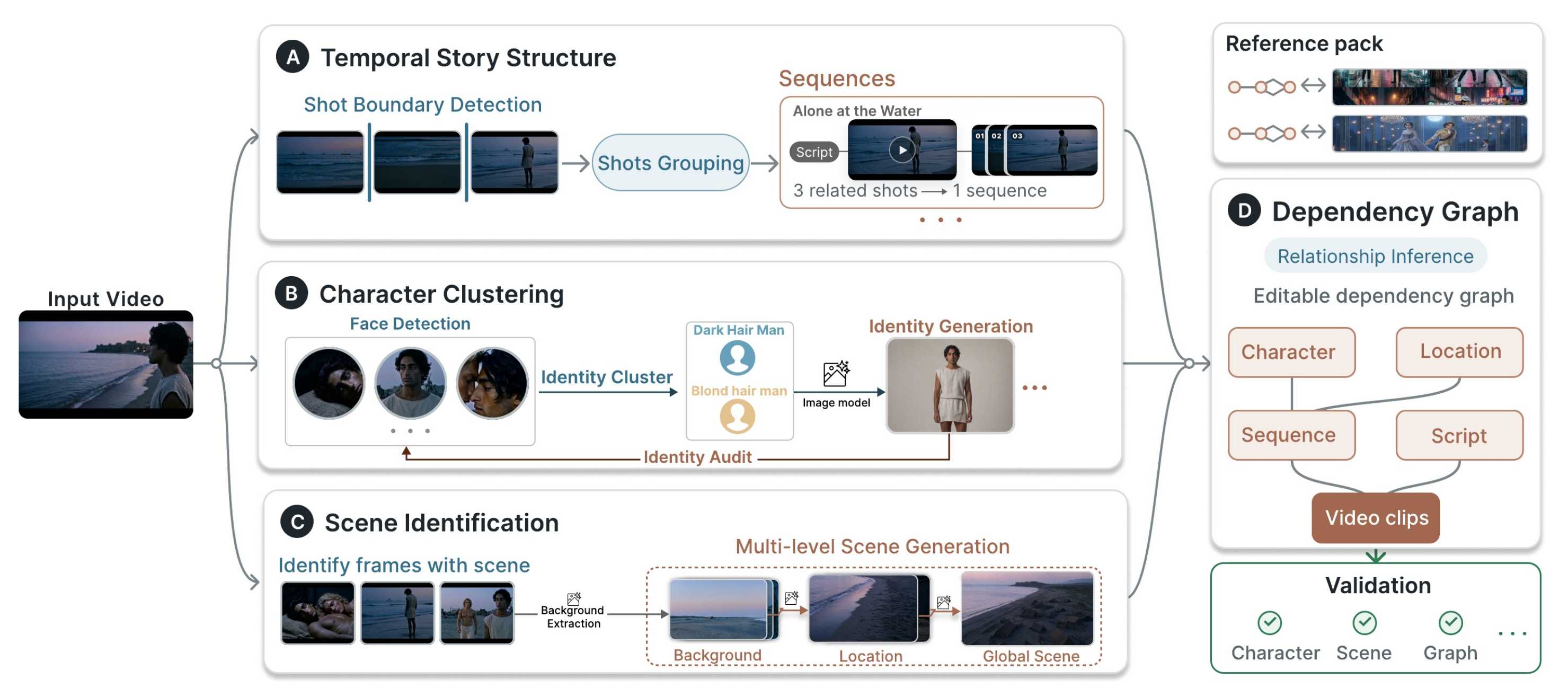}
    \caption{The pipeline for reconstructing the underlying video structure. (A) The video is first segmented into shots, which are grouped into sequences and used to infer the script with a VLM. (B) Faces are detected and clustered to identify characters. (C) Scenes are detected in a similar way across different levels, with support from reference packs. (D) The extracted elements and relationships are then organized into a hierarchical dependency graph, which is finally validated against the original video.}
    \label{fig:recover}
\end{figure*}

Here, we explain how \tool~ recovers the underlying elements and structure of an existing video(Fig.~\ref{fig:recover}). We design an agentic workflow that takes a video as input and generates a graph structure as output. The structure captures the elements and dependencies behind the video. One thing to note here is that this reconstruction may have multiple solutions. Although our approach is informed by how videos are typically created, our goal is not to recover the exact process used to produce the original video. Instead, we look for a plausible structure that can serve as a scaffold for users to understand and edit the video.

\subsubsection{Learning from forward creation traces.}
The agentic video creation workflow used in our formative study recorded a structured log for each generation run. Each log captures the process from the initial story to the final video, allowing us to pair the creation process (i.e., the underlying structure) with its resulting video.
We compile these pairs of data into a \textit{Process Reference Pack} and provide them to a Vision-Language Model (VLM) as in-context examples. The pack provides important guidance on how a video should be decomposed (e.g., how to identify the boundaries between different narrative events). The actual decomposition and reconstruction are performed by the agentic workflow described next.

\subsubsection{Temporal structure recovery}. To recover how a video unfolds over time, we first need to divide its continuous footage into meaningful temporal units. Camera cuts provide initial boundaries, but a single narrative event may span several adjacent shots. We therefore combine low-level shot detection with higher-level narrative grouping.
To divide the video into shots, we first use Mean Absolute Frame Difference to detect abrupt visual changes. The VLM uses selected key frames from each shot to infer the overall storyline and script, and merges adjacent shots that take place in the same scene or depict the same event into a higher-level narrative sequence. 

\subsubsection{Character identity recovery.}
We use MediaPipe~\cite{Lugaresi2019} to locate characters in keyframes and extract a crop of each person. A VLM analyzes these crops and groups appearances of the same person across shots, producing an initial identity group for each character. Using the grouped crops as references, the VLM then reviews the video to verify the groups and create dependencies. 
Finally, FLUX.2~\cite{blackforestlabs2025flux2} generates a clean identity reference image for each character using crops from multiple frames and viewpoints, together with an appearance description generated by the VLM.
The original crops retain the links between each character and the corresponding source shots.

\subsubsection{Scene hierarchy recovery.}
Because the same environment may appear across multiple shots with different viewpoints or lighting, we recover scene context at three levels. The agent first selects keyframes with clearly visible backgrounds. At the shot level, it removes characters from each keyframe to create a clean background plate. At the location level, the VLM groups adjacent shots that share the same environment, time, and event. At the global scene level, it links locations depicting the same underlying place despite differences in viewpoint or lighting, and uses their representative frames to construct a broader scene reference. 

\subsection{Guiding Video Editing with the Structure}
\label{sec:guide}
Once recovered, the video structure guides how an edit is interpreted, propagated, and executed. The agent translates a high-level editing intent into coordinated changes across the video.

\subsubsection{Interpreting an Editing Intent as a Global Instruction.}
\tool~ interprets each edit intent based on the user operation, the current workspace, and the video structure. The agent first determines the intended change and then traverses the dependency edges to identify the elements and shots it may affect. For each affected shot, it selects the necessary character images, scene references, and source clips, and specifies how they should be used. Providing only the references relevant to each shot reduces conflicting signals during generation.

\subsubsection{Turning a Global Instruction into Local Editing Tasks.}
Identifying the affected shots determines where an edit should be applied, but the instruction must still be adapted to each shot. For example, a general instruction such as ``replace the blond man'' may be ambiguous when several characters appear in the same frame. The VLM therefore inspects the original frame and describes the target in its local visual context, such as ``the blond man lying on the left side.'' The agent combines this description with the desired change and the content that should be preserved to produce an instruction for the visual generation model. The same process adapts other global instructions to the content of each affected shot.

\subsubsection{Execution.}
Once the instructions and visual references are prepared, \tool~executes them through a shared interface for reference-guided image and video editing. We use Qwen3.7 as our VLM agent, accessed through its API. We then use FLUX.2~\cite{blackforestlabs2025flux2} for image generation and editing. For video editing, the interface supports different models. On a local server with eight NVIDIA A100 GPUs, each with 80 GB of memory, we use JoyAI-Video-Edit~\cite{xiao2026joyaivideoedit} for rapid prototyping at approximately 16 FPS and Bernini~\cite{berniniteam2026bernini} for higher-quality generation, which takes approximately three minutes for a clip. The interface can also support commercial APIs such as Seedance~\cite{teamseedance2026seedance}.  
Finally, the agent uses the video structure to place each generated result in the corresponding shot and update the assembled video. Previous versions are retained in the version history, allowing users to review or revert changes at any time.
\section{Evaluation}

We evaluated \tool~ through two user studies. 1) We conducted a usability study with 12 participants spanning different levels of video-making experience, most of whom created videos for personal projects or social media. Through both controlled tasks and open-ended exploration, we examined the usability of the structure-based interface compared with the baselines.
2) We then conducted an expert review with five experienced filmmakers who had substantial experience in AI-assisted video production. Beyond usability, this review compared it with their previous practice and examined how \tool~ might fit into professional workflows, what creative opportunities it could introduce, and what improvements would be necessary for professional adoption. 


\subsection{Preliminary User Study}

\subsubsection{Participants.}
We recruited 12 participants (A1-A12) via social media and snowball sampling (6 male, 6 female, aged 22-34). We screened applicants based on information provided in their registration forms to ensure diversity in video creation backgrounds and experience. Based on their primary video-making practice, two participants were researchers working on AI video generation, two were social media content creators, and eight primarily created videos for personal projects. All participants had prior experience with AI-based video creation tools such as Seedance~\cite{teamseedance2026seedance} and Kling~\cite{kling}, and their overall video production experience ranged from three months to two years.

\subsubsection{Study Setup.}
We conducted the study in a hybrid format, with participants joining either in person or remotely. In-person participants used a macOS workstation with an Apple M2 Pro chip, while remote participants used their own computers. All participants connected to the same remote backend.
To compare \tool~with manual and automated workflows, we implemented two baselines. In the Manual Baseline, participants used CapCut to identify the shots relevant to the editing goal and divide the source video into clips of up to 30 seconds. We provided a separate interface for the video editing model, through which participants submitted each clip with a prompt and downloaded the edited result. They then returned to CapCut to assemble the clips into the final video. In the Agent Baseline, we equipped Codex with skills for video inspection, clip processing, generative editing, and video assembly. Participants provided the source video and editing goal through natural-language instructions, and Codex executed the complete workflow and returned an assembled video. All three conditions used the same video generation models and model configurations(described in Sec.~\ref{sec:guide}).

\subsubsection{Procedure.}
Each session lasted approximately 110 minutes. Before the study, participants reviewed and signed an informed consent form. Each participant received a US\$20 Amazon gift card as compensation.

\textit{Tutorial Session (15 minutes).}
Participants first received a brief introduction to the research goals and the system, including its overall pipeline and key features. We then collected participants personal information and details about their prior video creation experience. Using a prepared example, participants explored the recovered video structure and tried several editing operations to familiarize themselves with the interface. During this hands-on tutorial, we encouraged them to examine the different workspaces and features before proceeding to the study tasks.

\textit{Controlled Comparison Task (50 minutes).}
Each participant completed the same predefined editing task in three conditions: \tool, the Manual Baseline, and the Agent Baseline. For each condition, we provided the same prepared source video (about 3 mins) and editing goal(edit characters and scenes), and participants used the assigned workflow to produce an edited video. All conditions started from the original source video. The order in which participants began the three conditions was counterbalanced across participants to reduce learning effects. Because the Agent Baseline could take longer to complete, participants could proceed to another condition after submitting their instruction while Codex continued running in the background.

\textit{Open-Ended Exploration (30 minutes).}
Participants selected either their own video or one from a library we provided and then defined their own editing goal. They were encouraged to think aloud as they freely explored the system, tried different features, and edited different types of video elements.

\textit{Questionnaire and Interview (15 minutes).}
After completing the tasks, participants completed a questionnaire rating the system’s usability and their overall experience. They then took part in a brief semi-structured interview to explain their ratings and provide additional feedback on the system.

\begin{figure*}[t!]
    \centering
    \includegraphics[width=1\linewidth]{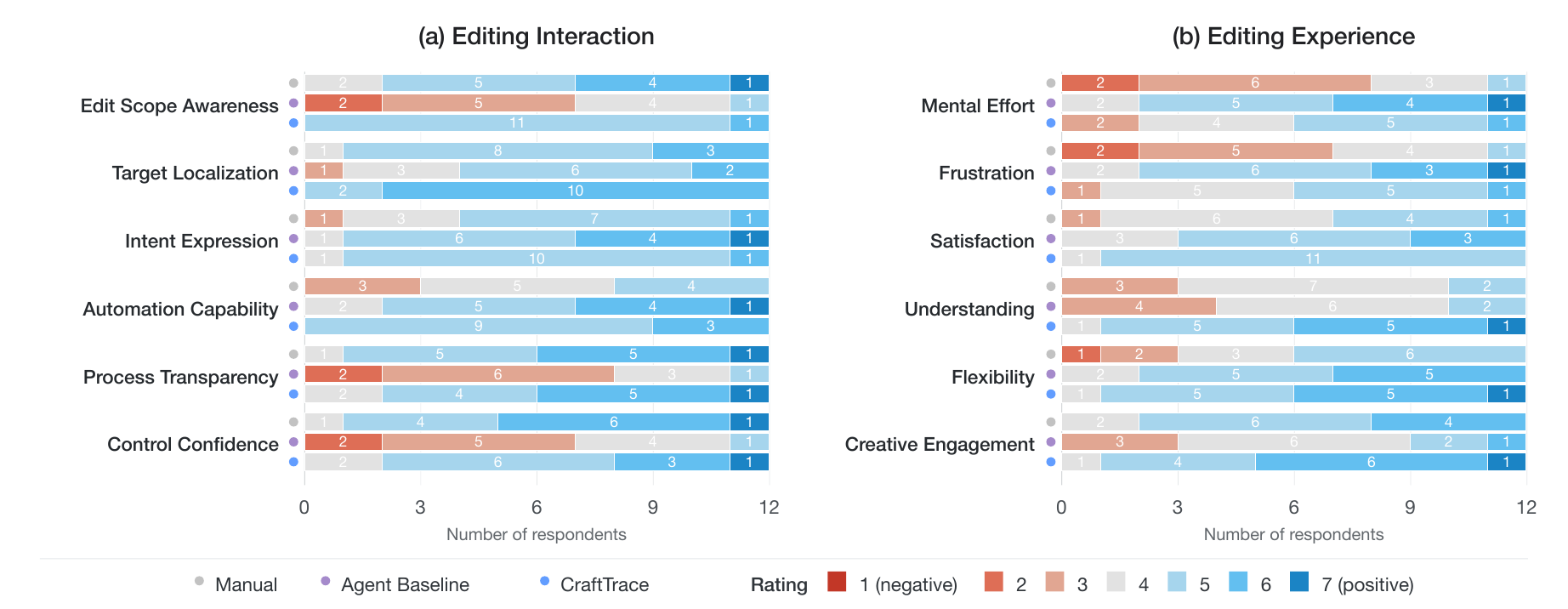}
    \caption{User ratings comparing the Manual Baseline, Agent Baseline, and CraftTrace across (a) editing interaction and (b) editing experience. Stacked bars show the distribution of seven-point ratings, with numbers indicating respondent counts. The higher is better. Given the current sample size, we focus on describing response distributions and do not conduct inferential statistical tests.}
    \label{fig:rating*}
\end{figure*}

\subsubsection{Results}
We combined quantitative and qualitative evidence to examine both how participants used the system and how they perceived it. Session recordings and system logs were used to characterize feature use and editing actions, while the generated videos were reviewed as the outcomes of these interactions. We summarized the Likert-scale ratings (Fig~\ref{fig:rating}) using descriptive statistics and used participants’ think-aloud comments and interview feedback to interpret the observed patterns.

\textit{User Preference.}
Participants consistently favored the structure-aware interaction provided by \tool. Most participants considered \tool~ is controllable (10/12), while 12/12 found it good for understanding the scope of an edit. 12/12 agreed that the workspaces helped them find their target efficiently, make it intuitive to express editing intent (11/12), and engage with the structure behind the video with editing automatically (10/12). A8 described the central benefit as \q{not having to watch the entire video and being able to see which nodes an element affects}.  As A7 noted, the fine-grained interaction supported local edits, although some details still needed to be \q{fixed in post-production}. Overall, participants valued \tool~ primarily for helping them understand, scope, and direct cross-shot edits, while generation fidelity remained a limitation.

\textit{User Strategies.}
Across the user study, we observed that participants developed diverse strategies for editing through the structure exposed by \tool. \textit{(1) The same editing intent could be expressed at different levels of intervention.} The workspaces are not isolated interfaces for mutually exclusive tasks, but provide complementary ways to work with the same underlying structure. For example, a character edit could be simply prompted in Graph or refined as a detailed attribute in Departments. Similarly, some participants revised a narrative direction in one step through Storyline, whereas others constructed the change incrementally in Graph by modifying the relevant script, character, scene nodes, and relationships. Participants selected among these paths according to their desired balance between convenience and control. A11 valued being able to \q{begin directly with an extracted element}, while A5 preferred finer-grained control for \q{repeated operations on component elements}. \textit{(2) Structural exploration helped some participants reshape their editing intent.} Rather than treating their initial goal as fixed, they revised it as they inspected the recovered elements and evaluated intermediate results. During open-ended exploration, A2 shifted from modifying a character’s overall appearance to edit the scenes related to the character. By surfacing possible points of intervention, the structural visualizations shaped what participants noticed and considered editable, allowing editing intent to emerge and evolve through interaction.
\begin{figure*}[h!]
    \centering
    \includegraphics[width=1\textwidth]{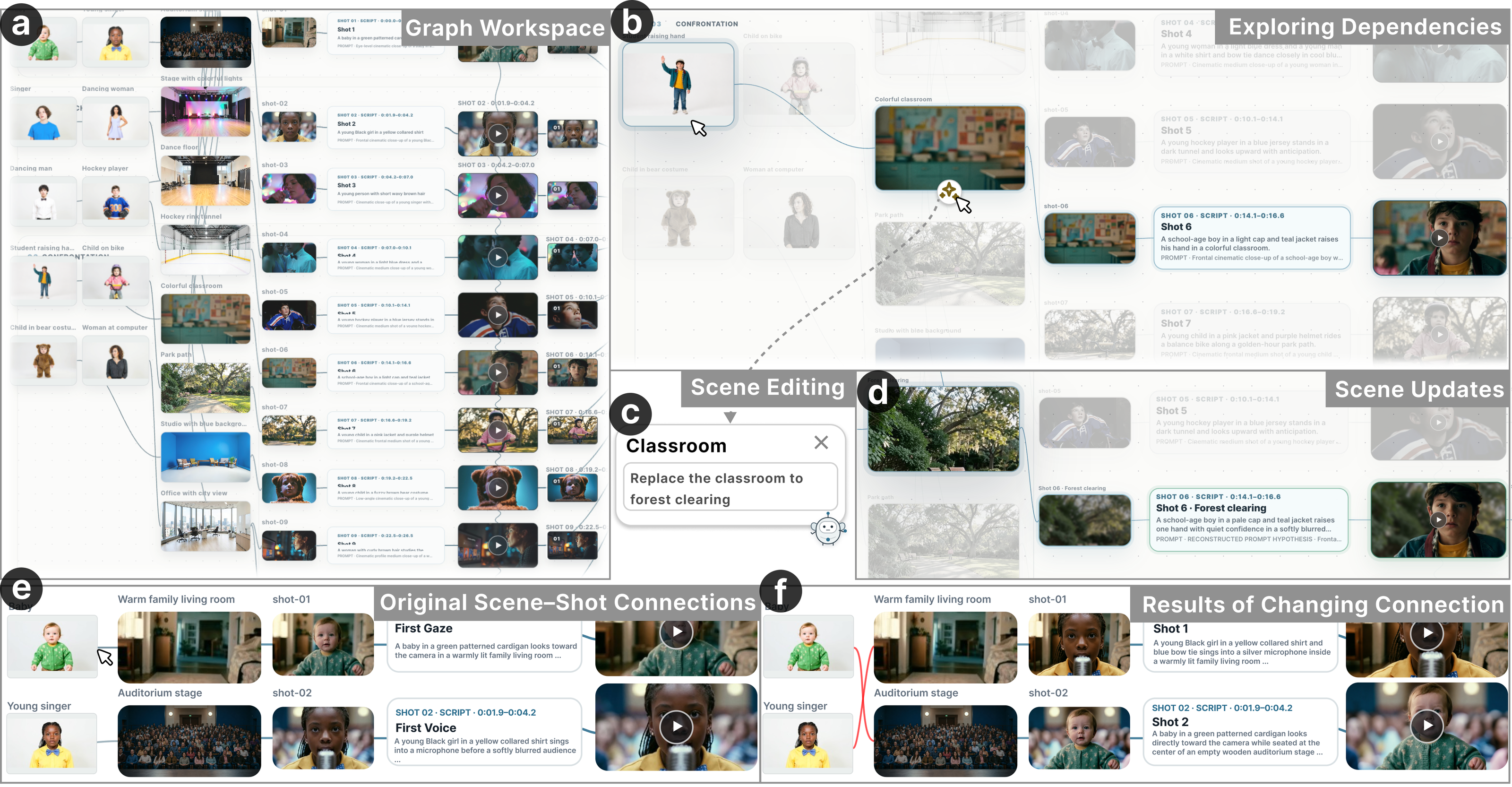}
    \caption{Example editing workflow performed by A3 on a montage-style short video. (a) \tool~ reconstructs the video structure, which A3 first examines in Graph mode. During the first edit, A3 (b) hovers over the student to reveal the linked scene and shot, (c) prompts \tool to replace the classroom with a forest scene, and (d) reviews the propagated updates to all affected scene, script, and shot nodes. During the second edit, A3 (e) examines the original character–scene mappings for Shots 1 and 2, then relinks the nodes to swap the two scenes. (f) \tool accordingly updates the scripts and regenerates the two affected shots.}
    \label{fig:case1}
\end{figure*}

\begin{figure*}[h!]
    \centering
    \includegraphics[width=1\textwidth]{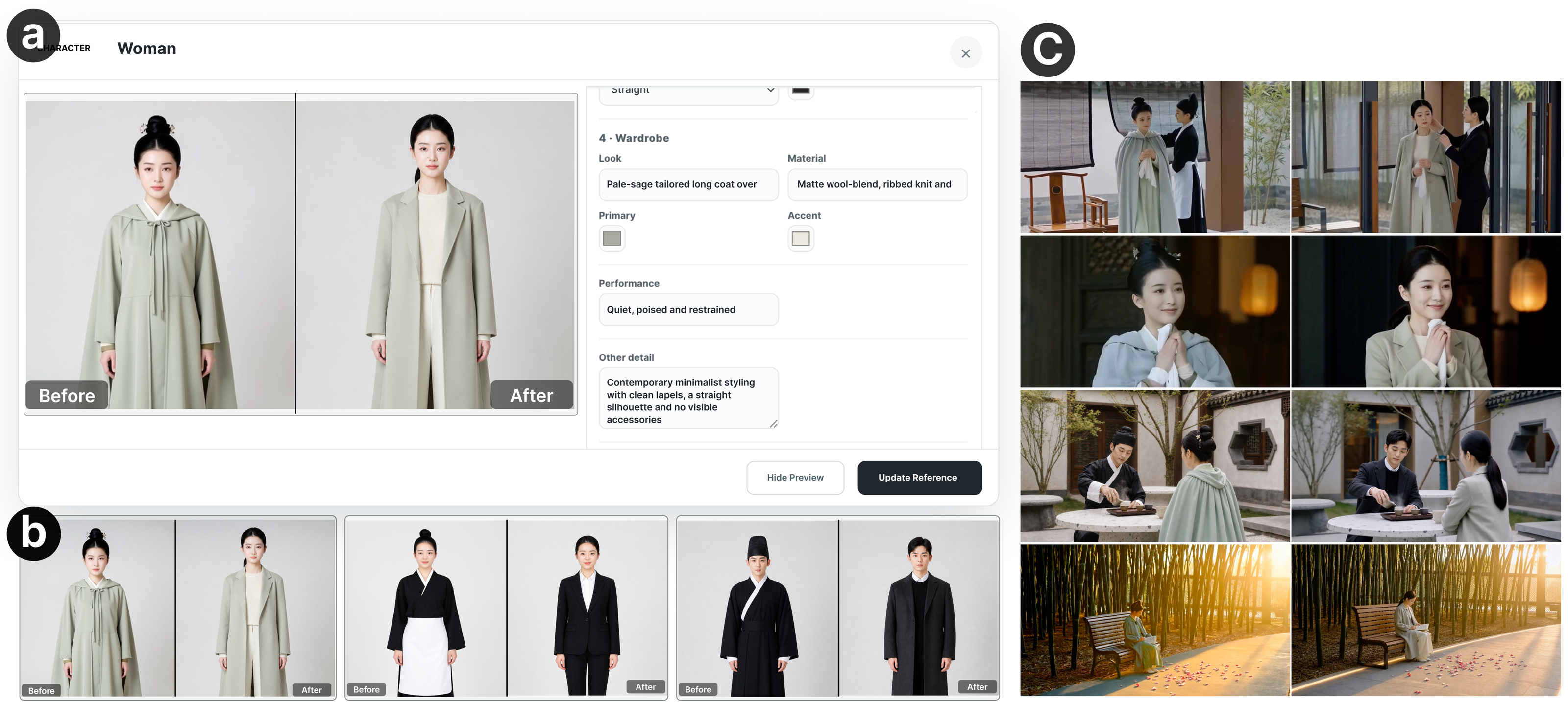}
    \caption{Example editing workflow performed by A8 on a traditional Chinese story. (a) In the \departmentsicon~Departments workspace, A8 edits the attributes of characters and scenes to give them a more modern appearance. (b) After iteratively adjusting the elements to the desired appearance, A8 confirms the edits. (c) \tool~ then applies these modifications and regenerates the entire video accordingly.} 
    \label{fig:case2}
\end{figure*}

\subsection{Expert Review}

\subsubsection{Participants.}
We recruited five film and media production professionals (E1--E5) through professional networks and industry contacts. They had an average of 17 years of professional experience. Their complementary expertise spanned directing and editing, visual effects and virtual production, character animation, sound design, and creative technology. All had hands-on experience incorporating generative AI into film or video production and had applied these tools across multiple major production projects. Their collective tool experience also included canvas-based generative authoring environments such as Adobe Firefly Boards and TapNow.

\subsubsection{Procedure}
While the preliminary user study focused on interactive experience with \tool, the expert review moved beyond usability to examine the broader conceptual and practical value, limitations, and future directions of the approach. We therefore did not include a comparative condition or tightly controlled tasks intended to measure interaction performance. The study was conducted remotely through video conferencing. We began with a 10-minute background interview about the expert's professional practice, experience with AI-assisted video creation, and generative video editing. We then provided a 15-minute walkthrough of the research motivation, design rationale, workflow, and key features of \tool. To ground the discussion in interactive experience, we gave the experts access to the online version of \tool~ and invited them to explore it freely for about 20 minutes. We then conducted a 40-minute semi-structured interview. We used a set of prepared questions to open the discussion and followed up on the observations that the experts comment from their own professional experience.

\subsubsection{Results.}The experts discussed both the opportunities enabled by structure-aware editing and the conditions required for integrating it into professional production. Below, we summarize their feedback.

\textit{\textbf{Organizing the video structure lays the groundwork for editing.}}
Expert feedback suggested that the recovered video structure is valuable even before creators commit to a specific edit. Creators do not always begin with a fully specified intent. They may first need to inspect the material, identify recurring elements, and determine the possible scope of a change. E1 described \q{helping you sort [and organize]} as the most valuable form of agentic assistance, envisioning a system that could retrieve \q{all related scene on one character show up} from a much larger collection. For professional creators, this kind of organizational and visual support may be more useful than fully automated editing: \q{maybe agent help a lot for novice, but for us, it could be better for just giving such kind of visualization}(E4). By reorganizing the video around meaningful elements and relationships, the recovered structure helps creators turn an initially vague idea into an actionable edit. The four workspaces further provide complementary views of the same video, corresponding to different aspects and stages of real-world filmmaking. Together, they shape the editing process by guiding what creators notice, what they regard as editable, and how they formulate the changes they want to make.

\textit{\textbf{The level of automation can be adapted to the editing context while the interface keeps users in control.}}
Expert feedback indicated that the appropriate level of automation depends on the stage of editing and the expected quality of the result. During ideation and previsualization, creators may accept greater agent autonomy because the goal is to explore possibilities quickly. E1 described generated alternatives as \q{a great starting point} from which creators could \q{pick and choose}, and viewed story branching as more suitable for \q{an ideation process as opposed to a post process}. As an edit moves toward final production, however, creators require closer supervision and have less tolerance for unintended decisions. The interface of structure provides a way to vary this level of control. By showing how an edit travels across related elements and shots(e.g., Fig~\ref{fig:graph}), \tool~ makes the agent’s interpretation, scope, and progress visible. Creators can review the proposed changes, restrict their scope, or intervene as the process unfolds. Automation in \tool~ becomes an inspectable and adjustable process rather than a single black-box action, allowing creators to grant the agent more freedom during exploration while retaining tighter control for production-oriented editing.

\textit{\textbf{Full video editing depends on knowing what not to change.}}
A successful full video edit requires not only modifying the intended clips correctly but also preserving all unrelated footage. As E4 explained, \q{With an end-to-end video editing model, you may input a long, multi-shot video even though only a few shots contain the target you want to edit} (E4). Experts therefore regard the preservation of original frames not merely as a matter of visual quality, but also as a \q{production management requirement} (E3). In professional workflows, a completed shot may already embody substantial creative and technical effort, leaving little tolerance for unintended changes introduced by a generative model. Current models such as Seedance~\cite{teamseedance2026seedance} support editing multi-shot videos of up to 30 seconds. However, when only a few shots require modification, diffusion-based regeneration may still introduce subtle variations in otherwise unaffected footage. In contrast, \tool~ makes the scope of AI intervention explicit: \q{we clearly know where the AI model has been applied throughout the video, which is reassuring because we do not need to carefully inspect every unaffected shot again}(E1).

\textbf{Recovered structure preserves the project behind the final video.} Experts emphasized that a finished video should not mark the end of a project, as its prompts, references, and creative decisions may still be needed for future revisions. E5 described a previous need to migrate a video project created with an earlier Kling~\cite{kling} model to the more advanced Seedance 2.5, a task that would be \q{really easy} with \tool. By recovering a reusable structure, \tool~could preserve the project's established intent while allowing it to be reopened with new possibility. The same structure could keep project knowledge accessible across tools and collaborators. Rather than reconstructing the exact production history, \tool~could turn a closed video into persistent project memory, which is a working state that can be searched, revised, migrated, and shared. As E1 mentioned, \q{\tool~is like reconstructing everything we built step by step. When we give it an old video project, we can pick up where we left off almost immediately, apply changes across the project automatically, and still control everything through the interface.}

\textit{\textbf{Advances in end-to-end editing do not diminish the value of structure-based interactivity.}}
When discussing increasingly end-to-end workflows, we asked experts where \tool~ would remain valuable. For example, a general-purpose agent such as Codex could locate relevant shots, divide the footage, submit each segment to video-editing models, and assemble the final result. Alternatively, future generative video models may directly process videos lasting several minutes rather than being limited to 30 seconds. Expert (E4, E5) argued that the central value of \tool~ lies in externalizing the structure. End-to-end pipelines may simplify execution, but they often compress a complex editing process into a single prompt and output, leaving users with limited awareness of \q{how the video is organized}, \q{how the agent interprets their intent}, or \q{how generative editing is carried out across the video}(E4). E2 similarly emphasized the value of letting users \q{directly manipulate} the structure to express their editing intent. The structure also serves as \q{reusable context shared by the user and the agent}(E1). Instead of requiring the agent to analyze the video from scratch for every new request, subsequent edits can build more efficiently on the same explicit representation. The value of structure-based interaction therefore lies in making generative video editing more transparent, efficient, and controllable.
\section{Discussion}

\subsection{Implicit and explicit memory for generative video editing}
In a from-scratch agentic video-creation process~\cite{li2024animdirector}, a memory bank maintains characters, scenes, and narrative context~\cite{zhang2025storymemmultishotlongvideo}. \tool~can be viewed as its inverse: it reconstructs this context from a finished video as an interactive structure for inspection and editing. This provides explicit semantic memory of elements and relationships, but visual continuity also depends on implicit generative memory, such as KV-Cache in latent space~\cite{yang2025longliverealtimeinteractivelong}, to preserve appearance and motion. Our current system primarily manages the former.

This limitation becomes apparent when edits span multiple generation requests. \tool~normally edits each shot separately, but continuous shots longer than 30 seconds exceed our current model's input duration and must be divided into shorter segments. Conditioning each segment on the previous segment's final frame helps align appearance at the boundary, but changes in motion speed or rhythm may remain visible. Across related shots, separate requests may also produce differences in facial details, texture, or lighting despite using the same character and scene references. Such variation is common in manual workflows, where creators often regenerate clips repeatedly to obtain sufficiently consistent results (E1).

Future work could combine structure-aware agentic editing with latent state management, allowing the agent to determine when implicit states should be carried forward, selectively reused, or reset. KV caches~\cite{yang2025longliverealtimeinteractivelong} or other temporal states could persist across segments of a continuous shot to help preserve motion continuity, while identity and style context could be retained across connected shots as motion and viewpoint change.

\subsection{Evolving states of elements}
\tool~extracts recurring elements, such as characters and scenes, as reusable visual anchors. The agent follows the structure to provide these anchors as references for affected shots, helping maintain subject identity and scene appearance as edits propagate~\cite{zhou2024storydiffusion,kara2025shotadapter}. However, the same element may take different states throughout a narrative, making a fixed reference less suitable.

An element may evolve over time (e.g., a character may acquire a scar, or a location may change after objects are moved). Reusing the same anchor across these changes may suppress intended variation or reintroduce an outdated state. A narrative may also intercut different periods (e.g., in a montage showing the same character at different ages). In this case, the appropriate reference depends on the period depicted in each shot, rather than its position in the video. This also poses a challenge in the Storyline workspace, where revising the narrative may also change which state should appear, particularly when shifts between periods overlap with changes in the element itself.

Future work could therefore extend structure recovery and editing to represent both recurring identities and their changing states. Each identity could be associated with multiple states, linked to the narrative events and shots in which they apply. Users could inspect and edit these states and their transitions, adjust their temporal or narrative scope, and control how changes propagate. For nonlinear narratives, the structure would also need to maintain different states across story branches, allowing references to follow the narrative context of each shot.

\subsection{Applications Across Video-Creation Stages}
Our studies suggest that \tool~could support both ideation and post-production, with different workflow requirements. Ideation prioritizes rapid exploration and may tolerate approximate results, favoring faster generation and greater agent autonomy, making models such as JoyAI~\cite{xiao2026joyaivideoedit} a suitable fit. Post-production emphasizes output quality, preservation of the original footage, and careful review of propagated changes, making Bernini~\cite{berniniteam2026bernini} or Seedance~\cite{teamseedance2026seedance} more suitable. Because \tool~is agnostic to the underlying model, implementations could adapt model choice, automation, and confirmation steps to the creator's current stage.

During ideation, creators could reconstruct an existing video into our graph structure to examine its shot design and narrative pacing, then replace its cast or setting to preview their own concept. The Storyline workspace could support alternative story directions, while the Departments workspace could support character and environment designs. These results serve as previsualizations before committing to production.

In post-production, the recovered structure could turn a finished video back into a reusable project. One expert wanted to recreate an earlier project using a newer generation model. Recovering its elements and shot relationships could coordinate regeneration without rebuilding the workflow from scratch. Similarly, localization could adapt an AI short drama for an Asian audience with Western characters and settings while preserving its storyline, performances, framing, and pacing. Creators could express this as a high-level change and use the structure to update relevant footage rather than revise every shot individually.

Practical applications should also connect high-level, cross-shot editing with iterative refinement of individual shots. Creators still need to regenerate shots, compare alternatives, and correct local artifacts. Future interfaces could support candidate histories, local editing, and locking approved clips, allowing refinement without affecting completed parts of the video~\cite{TapNow_2026,Guo_2026_Protosampling}.

\subsection{Scalability and Granularity of the Recovered Structure}

\textit{Scalability to longer videos.}
Our evaluation focused on videos up to a few minutes long, reflecting the short-form AI video creation scenarios considered here. In principle, the workflow is not tied to a fixed video duration, as \tool~coordinates repeated calls to video models to edit relevant shots. However, longer videos would produce larger structures and increase waiting time for edits affecting many shots, limiting rapid iteration. Supporting such videos therefore requires interfaces with different levels of detail, allowing users to navigate an overview and inspect individual elements and shots when needed. Further evaluation is required to assess scalability and usability beyond the tested durations.

\textit{Control over finer details.}
Professional workflows may require finer control than the current structure provides. Experts identified additional elements and attributes, including props, lighting conditions, camera angles, color schemes, and details such as a logo on a wall. Representing these would support more precise edits but increase reconstruction uncertainty and the effort needed to inspect and manage the structure. Future systems could recover additional details according to the editing task and provide references suited to those details. For example, 3D models or 360-degree images could serve as spatial proxies beyond image and video references. Integration with generative authoring tools such as TapNow~\cite{TapNow_2026} could also let creators generate or refine reference assets within existing workflows and use them to guide edits across the video.
\section{Conclusion}

We introduced \tool~, a prototype interface that reconstructs a plausible structure behind a finished video and uses it as a scaffold for user-driven, agent-assisted editing. \tool~ presents the same underlying structure through task-centric workspaces, allowing users to approach the video from different perspectives. Users can modify an individual element and propagate the change across its dependent shots, or modify the relationships between elements to reshape the video structure. Our evaluations suggest that \tool~ supports full multi-shot video editing by reducing the effort required to locate affected shots and coordinate changes across them, particularly during rapid exploration and brainstorming. More broadly, this work frames generative video editing as the manipulation of an editable structure that coordinates changes across clips. By making the structure behind a video visible, malleable, and actionable, \tool~ provides common ground through which users can express creative intent and AI agents can coordinate its realization across the video.


\bibliographystyle{ACM-Reference-Format}
\bibliography{reference}

\clearpage
\appendix
\section{Runtime Comparison}
\label{app:technical-evaluation}

We compared the runtime of \tool{} and Codex to assess the efficiency gains of our recovery workflow and editing with a recovered structure over a general-purpose agent workflow. Both conditions used the same image and video generation models deployed on a local server, with JoyAI-Video-Edit selected for its fast generation speed. By holding the generation backend constant, we focused this evaluation on workflow runtime rather than output quality. Because informal visual inspection in our test suggested broadly comparable output quality since we use the same generation model.

\paragraph{Recovery.}
We evaluated 20 videos longer than 30 seconds, with a mean duration of 118 seconds. Both conditions started from identical source videos and targeted a complete editable structure. Codex was equipped with a recovery skill specifying the same required outputs, including temporal structure, character and scene references, scripts, and dependency relations. Under these conditions, \tool{} completed recovery in an average of 5.59 minutes, compared with 29.88 minutes for Codex.

\paragraph{Editing.}
We evaluated 50 editing tasks across 10 videos, with five tasks per video. Both conditions received identical source videos, instructions, and target annotations. \tool{} reused the recovered structure, whereas Codex independently inspected the video and planned the edit, with access to records created within its own session. We measured editing workflow time excluding prior recovery and waiting for media generation and queueing. The comparison showed a shorter average editing time for \tool{} (2.99 minutes) than for Codex (10.47 minutes).

\section{User Study Questionnaire}
\label{app:questionnaire}

The questionnaire covers editing interaction and editing experience. In the statements below, ``this workflow'' refers to the condition being evaluated: the Manual Baseline, the Agent Baseline, or \tool. All items use a seven-point Likert scale, ranging from 1 (strongly disagree) to 7 (strongly agree), with 4 indicating neither agreement nor disagreement. Higher ratings indicate more favorable experiences for all items.

\subsection{Editing Interaction}

\begin{enumerate}
    \item \textbf{Edit Scope Awareness.}
    I could tell which parts of the video would be affected by my edit.

    \item \textbf{Target Localization.}
    It was easy to find the shots relevant to my editing goal.

    \item \textbf{Intent Expression.}
    I could clearly express the changes I wanted to make using this workflow.

    \item \textbf{Automation Capability.}
    This workflow automated the steps needed to carry out my edit.

    \item \textbf{Process Transparency.}
    I could follow how my edit was carried out.

    \item \textbf{Control Confidence.}
    I felt confident that I could control how my edit was applied.
\end{enumerate}

\subsection{Editing Experience}

\begin{enumerate}
    \item \textbf{Mental Effort.}
    Completing the task with this workflow required little mental effort.

    \item \textbf{Frustration.}
    I experienced little frustration while using this workflow.

    \item \textbf{Satisfaction.}
    I was satisfied with my overall experience using this workflow.

    \item \textbf{Understanding.}
    This workflow helped me understand how different parts of the video were related.

    \item \textbf{Flexibility.}
    I could adjust my editing approach as my goals changed.

    \item \textbf{Creative Engagement.}
    I felt actively involved in making creative decisions while using this workflow.
\end{enumerate}
\section{Expert Review Participants}

\begin{table*}[th!]
\centering
\small
\renewcommand{\arraystretch}{1.15}
\begin{tabularx}{\textwidth}{@{}lXccc@{}}
\toprule
Expert & Role
& \shortstack{Video creation\\frequency}
& \shortstack{Video GenAI use\\frequency}
& \shortstack{Video creation\\experience (years)} \\
\midrule
E1 & Filmmaker and creative technologist
& Daily  & Daily  & $>15$ \\
E2 & Composer and sound designer
& Daily  & Daily  & $>10$ \\
E3 & 3D generalist and character technical director
& Weekly  & Weekly  & $>20$ \\
E4 & CG artist, supervisor, and educator
& Weekly  & Daily  & $>10$ \\
E5 & AI filmmaker and performer
& Daily  & Weekly  & $>20$ \\
\bottomrule
\end{tabularx}
\caption{Profiles of the five experts.}
\label{tab:expert-profiles}
\end{table*}

\section{Duration Limits and Editing Constraints of Video Models}
We summarize the duration limits and editing constraints of several video models that we tested or used in this work.

\begin{itemize}
    \item \textbf{Seedance.} Seedance 2.0 can generate videos of up to 15 seconds in a single pass, whereas Seedance 2.5 extends this duration to 30 seconds and supports repeated extensions. 

    \item \textbf{Bernini.} The official inference interface defaults to 81 frames at 16 FPS, corresponding to approximately five seconds. 

    \item \textbf{Aleph 2.0.} Runway Edit Studio supports editing both individual shots and sequences containing multiple shots, up to 30 seconds in length. 

    \item \textbf{JoyAI-Video-Edit.} JoyAI supports causal streaming editing without requiring a predefined total duration. However, in our tests, applying the same instruction across multiple shots could introduce unintended changes in shots unrelated to the intended edit. Therefore, to use the model in a multi-shot full video, we need to identify and extract the relevant clips and provide a prompt based on the content of each clip.
\end{itemize}

\end{document}